\documentclass[a4paper,11pt]{article}
\pdfoutput=1 

\usepackage{jheppub} 
\usepackage{mathrsfs,graphicx,rotating,amsmath,amsfonts,mathtools,booktabs,amssymb,wasysym,caption}
\usepackage{hyperref}
\usepackage{slashed}
\usepackage{natbib}
\usepackage[table,xcdraw,dvipsnames]{xcolor}
\usepackage{graphicx}
\usepackage{subfig}
\usepackage{bbold}
\usepackage[utf8x]{inputenc}
\usepackage[english]{babel}
\usepackage{multirow,multicol}
\usepackage{epstopdf}
\usepackage{bbm}
\usepackage{appendix}
\usepackage{libertine}
\usepackage{braket}
\usepackage{wasysym}
\usepackage{empheq}
\usepackage{cancel}
\usepackage{enumitem}
\usepackage{mathrsfs}
\usepackage{lmodern}
\usepackage{tabularx}
\usepackage{multicol}
\usepackage{color}
\usepackage{mathtools}
\usepackage{placeins}
\usepackage[abs]{overpic}

\DeclareMathOperator{\e}{e}

\DeclareMathOperator{\Tr}{Tr}

\renewcommand{\d}{{\rm d}}
\newcommand{\arctanh}{{\rm arctanh}}
\newcommand{\U}{{\mathrm{U}}}
\renewcommand\({\left(}
\renewcommand\){\right)}
\renewcommand\[{\left[}
\renewcommand\]{\right]}
\newcommand{\Vol}{{\rm Vol_{4}}}
\renewcommand{\U}{{\rm U}}
\newcommand{\UoneV}{{\U(1)_{V}}}
\newcommand{\SU}{{\rm SU}}
\newcommand{\A}{{\mathsf A}}
\newcommand{\B}{{\mathsf B}}
\newcommand{\0}{{\mathsf 0}}
\newcommand{\1}{{\mathsf 1}}

\title{\boldmath Fermions at finite density in the path integral approach}

\author[a]{Alessandro Podo,}
\author[b]{Luca Santoni}

\affiliation[a]{Department of Physics, Center for Theoretical Physics, Columbia University, New York, 538 West 120th Street, NY 10027, USA}
\affiliation[b]{Universit\'e Paris Cit\'e, CNRS, Astroparticule et Cosmologie, 10 Rue Alice Domon et L\'eonie Duquet, F-75013 Paris, France}

\emailAdd{ap3964@columbia.edu}
\emailAdd{santoni@apc.in2p3.fr}

\abstract{
We study relativistic fermionic systems in $3+1$ spacetime dimensions at finite chemical potential and zero temperature, from a path-integral point of view. We show how to properly account for the $i\varepsilon$ term that projects on the finite density ground state, and compute the path integral analytically for free fermions in homogeneous external backgrounds, using complex analysis techniques. As an application, we show that the $\U(1)$ symmetry is always linearly realized for free fermions at finite charge density, differently from scalars. We study various aspects of finite density QED in a homogeneous magnetic background. We compute the free energy density, non-perturbatively in the electromagnetic coupling and the external magnetic field, obtaining the finite density generalization of classic results of Euler--Heisenberg and Schwinger. We also obtain analytically the magnetic susceptibility of a relativistic Fermi gas at finite density, reproducing the de Haas--van Alphen effect. Finally, we consider a (generalized) Gross--Neveu model for $N$ interacting fermions at finite density. We compute its non-perturbative effective potential in the large-$N$ limit, and discuss the fate of the $\U(1)$ vector and $\mathbb{Z}_2^A$ axial symmetries.
}

\begin{document}
\maketitle
\flushbottom

\section{Introduction}

It has been long appreciated that systems of free scalars and free spin-$1/2$ particles at low temperature and nonzero charge density have strikingly different properties. The former give rise to the phenomenon of Bose--Einstein condensation and are characterized by the spontaneous breaking of an internal symmetry. The latter fill out a Fermi sphere in momentum space, and have unbroken internal symmetry. These properties can be traced back to the statistical properties of the underlying  fields, bosonic and fermionic respectively, as dictated by the spin-statistics theorem in relativistic quantum field theory (QFT). 

In the presence of interactions, the fate of the internal $\U(1)$ symmetry in a state of finite charge density is less clear. For systems of interacting scalar fields, in~\cite{Nicolis:2023pye} we have provided evidence that the ground state at finite chemical potential cannot develop a charge density unless it spontaneously breaks the $\U(1)$ symmetry (see also~\cite{Nicolis:2011pv,Joyce:2022ydd} for previous related discussions). We have proven this statement in perturbation theory at one loop for generic non-derivative scalar's self-interactions, and in a ${\rm O}(N)$ vector model at large $N$. A complete proof is still missing, but these results are suggestive of the generality of the phenomenon. {This connection plays a crucial role also in the superfluid effective theory approach to the large charge sector of Conformal Field Theories~\cite{Hellerman:2015nra,Monin:2016jmo,Jafferis:2017zna,Cuomo:2022kio} (see also~\cite{Gaume:2020bmp} for a review). An important alternative phase is represented by Fermi liquids~\cite{Alberte:2020eil,Komargodski:2021zzy,Dondi:2022zna} and one would like to understand under what conditions this phase arises.}

A difficulty that one encounters when trying to address this question in full generality is that of properly accounting for the statistics of the fields~\cite{Nicolis:2023pye}. This should enter non-trivially to encode the different properties of bosons and fermions at nonzero charge density. In order to make progress in this direction, it seems instructive to reconsider the question about spontaneous symmetry breaking in fermionic QFTs at finite chemical potential and analyze the differences with respect to bosonic systems. In this spirit, our goal in this work is to revisit for fermionic fields some of the aspects that we addressed in~\cite{Nicolis:2023pye} for scalar theories. In particular, we will pose the following question: \emph{given a fermionic theory at finite charge density, is the $\UoneV$ symmetry realized linearly or spontaneously broken?} To answer the question, we will adopt a path-integral approach and  show explicitly that, to correctly compute physical  quantities such as the free energy density of a relativistic  Fermi gas, one needs to carefully take into account the $i\varepsilon$ term in the path integral---which allows to  project on the correct ground state of the Hamiltonian at finite $\mu$. As we will discuss, the $i\varepsilon$ term is a key ingredient that is responsible for the different analytic structure in the path integral and the resulting partition functions of fermions and bosons.  
Analyses of systems at finite chemical potential (or density) and zero temperature are often formulated as the $T\to 0$ limit of finite temperature QFT calculations.
The subtlelties associated to the zero-temperature limit have been analyzed in an interesting recent work~\cite{Gorda:2022yex}. Motivations include the study of the phase diagram of QCD at finite density~\cite{Alford:2007xm,Fukushima:2010bq,Kurkela:2009gj}. In this article we take a complementary approach, and discuss how to consistently perform finite density calculations in fermionic QFTs directly at $T=0$.
With this understanding, we will compute the partition function of a free Fermi gas in the presence of a source term in the path integral---which will play the role of an order parameter for the $\UoneV$ symmetry---and show that the vacuum expectation value of the order parameter vanishes in the limit in which the source goes to zero, proving that the $\UoneV$ symmetry is unbroken for free fermions. We will then extend the simple case of the free Fermi gas in two different directions. 

First, we will include a non-trivial magnetic background for the Fermi gas and compute explicitly the fermionic functional determinant in the presence of a magnetic field. 
Our results can have applications in the study of QED in intense magnetic backgrounds and at finite density. The dynamics of QED with strong external fields is an old subject~\cite{Heisenberg:1936nmg,Weisskopf:1936hya,Schwinger:1951nm}, see e.g.~\cite{Hattori:2023egw} for a recent review. Applications range from the astrophysics of strongly magnetized pulsars~\cite{Harding:2006qn,Marklund:2006my,Kim:2021kif} to laser and plasma physics~\cite{Fedotov:2022ely,Gonoskov:2021hwf}.
The quantum effective action of QED at finite temperature and density has been computed in Refs.~\cite{Elmfors:1993wj,Elmfors:1993bm,Sharapov:2003te}. Similar results have been obtained in $2+1$ dimensions~\cite{Gusynin:1994va}. These works considered systems at finite temperature and express the results as thermal integrals. With our approach, working directly at $T=0$, we obtain closed form analytic results, non-perturbatively in the electromagnetic coupling and magnetic field.
As a warm up, at zero chemical potential, we provide a modern derivation of the Euler-Heisenberg quantum effective action in its zeta function representation~\cite{Dittrich:1985yb,Blau:1988iz,Dunne:2004nc}.
We then obtain closed form analytic expressions for the finite density contributions, including the de Haas--van Alphen effect.

Second, we will include interactions among fermions. We will do so by considering a $3+1$ dimensional generalization of the Gross--Neveu model, for a system of $N$ Dirac fermions transforming in the fundamental representation of a vectorial ${\rm U}(N)$ global symmetry group. In the limit of large $N$, the model is solvable analytically and allows one to compute the fermionic path integral exactly. Although we will not prove in full generality that the $\UoneV$ symmetry is unbroken at finite density, we will provide evidence that the system can support a finite density phase with unbroken symmetry. This should be contrasted with the result discussed in~\cite{Nicolis:2023pye} for the ${\rm O}(N)$ vector model, where we showed that, for a system of $N$ scalars, the breaking of the $\U(1)$ symmetry is instead inevitable at nonzero charge density, in the large-$N$ limit.

The work is organized as follows. In section~\ref{sec:SSB} we start with some general considerations about the spontaneous breaking of the $\UoneV$ symmetry in fermionic theories at finite density from a path-integral point of view. In section~\ref{sec:fermionsfinitemu} we discuss the role of the $i\varepsilon$ term in the fermionic path integral, showing the difference with respect to the bosonic case, and provide a proof that the $\UoneV$ symmetry is unbroken for free fermions. In section~\ref{sec:magnetic}, we discuss various aspects of a Fermi gas in a magnetic background, including a derivation of the QED effective action and magnetic susceptibility at finite density.  The (generalized) Gross--Neveu model is then discussed in section~\ref{sec:interacting}. Some technical aspects and useful relations are collected in the appendices. 

\vspace{.2cm}

\noindent
\textbf{Conventions:}
We adopt natural units $\hbar=c=1$, and the mostly minus convention for the Minkowski spacetime metric. We assume that the chemical potential $\mu$ is non-negative ($\mu\geq0$), for ease of notation.
The free energy density as a function of $\mu$, is defined as 
\begin{equation}
f(\mu)\equiv(-i \, \Vol)^{-1}\left(\log \dfrac{Z\[\mu\] }{Z\[0\] }\right).
\end{equation}
To simplify the notation we will suppress factors of volume. When needed, they can be reintroduced by dimensional analysis. Conventions on $\gamma$ matrices are detailed in appendix~\ref{app:conventions}. The symbol $\bar{\mu}$ is reserved for the $\overline{\rm MS}$ renormalization scale, and its occurrence should be clear by context.

\section{Symmetry breaking at finite density?}
\label{sec:SSB}

In order to address the question of whether  the $\UoneV$ symmetry is linearly realized  or spontaneously broken in a fermionic theory at finite density, we look for an order parameter for the symmetry, that is: an operator $O$ which has a non-vanishing commutator with the charge operator
\begin{equation}
 \delta \hat O = \[ \hat Q, \hat O \] \neq 0.
 \end{equation}
If such an operator has non-zero expectation value on the ground state, i.e.~$\langle \delta O \rangle\neq0$, the symmetry is \emph{spontaneously broken} (i.e.~the ground state is not invariant under $Q$).

Let us start by considering a system of \emph{free} massive Dirac fermions in $3+1$ dimensions at finite density. The underlying dynamics is controlled by the Lagrangian
\begin{equation}\label{eq:Dirac_free}
\mathcal{L} = \bar{\Psi} \( i \slashed{\partial} - m \) \Psi.
\end{equation}
This theory has a continuous $\rm U(1)$ symmetry
\begin{equation}
{\rm U(1)_{V}}:
  \begin{cases}
  \;  \Psi \longrightarrow e^{i \alpha} \, \Psi \;,\\
  \;  \bar{\Psi} \longrightarrow e^{- i \alpha} \, \bar{\Psi} \;.
  \end{cases}
\end{equation}
The Noether current associated to the symmetry is
\begin{equation}
 j^{\mu}_{V} = \bar{\Psi} \gamma^{\mu} \Psi,
 \end{equation} 
corresponding to the number of left-handed plus right-handed particles (the vectorial current).

We would like to show that all the (local) order parameters quadratic in the field variables have zero expectation value, suggesting that the symmetry is linearly realized. The argument we present here is naive, since it \emph{assumes} that the partition function is differentiable with respect to the Majorana sources $j$ and $j^*$ around $(j,j^*)=(0,0)$. We will justify this assumption later in explicit examples of fermionic theories.

A prototypical order parameter for the vectorial $\UoneV$ symmetry is the Majorana term 
\begin{equation}
\delta O_{\rm Maj} = \overline{\Psi^{c}} \;\Psi = +i \,\Psi^{T}(\gamma^{2})^{*}\gamma^{0} \,\Psi .\end{equation}
In order to compute its vacuum expectation value we use the path integral formulation for fermions at finite $\mu$ and include a source term. The finite $\mu$ action will be rederived later, but for now let us consider directly the finite $\mu$ path integral with Majorana source
\begin{equation}
Z\[j;\mu\]= \int D\bar{\Psi} D\Psi \, \exp\[ i \int \d^{4}x \, \bar{\Psi }\( i \slashed{\partial} + \gamma^{0} \mu - m \) \Psi + j\frac{1}{2} \,\Psi^{T}(\gamma^{2})^{*}\gamma^{0} \,\Psi + \rm h.c. \].
\end{equation}
The vacuum expectation value of the Majorana order parameter is given by
\begin{equation}
\label{vev}
\langle \delta \hat O_{\rm Maj}\rangle_{\mu} = \dfrac{1}{Z_{\mu}}\dfrac{\delta Z_{\mu}}{\delta j } \Bigg\vert_{j=0} = \dfrac{\delta \log Z_{\mu}}{\delta j } \Bigg\vert_{j=0}.
\end{equation}
Since the action is quadratic in the field variables, the path integral can be performed exactly and expressed as a determinant. The finite $\mu$ Lagrangian, including source terms, can be written as follows (up to boundary terms):
\begin{equation}
\label{quad}
\mathcal{L}_{\mu}\[j\] = - \dfrac{i}{2} 
\begin{pmatrix}
\Psi  &
\bar{\Psi}
\end{pmatrix}  
\begin{pmatrix}
\B & \A \\
-\A & \B^{\dagger}
\end{pmatrix}
\begin{pmatrix}
\Psi  \\
\bar{\Psi} 
\end{pmatrix},
\end{equation}
where $\A = [ -i \slashed{\partial}_{x} - \gamma^{0} \mu + m ]_{\alpha\beta} \, \delta^{4}(x-y)$ and $\B = -j \, [(\gamma^{2})^{*}\gamma^{0}]_{\alpha\beta} \, \delta^{4}(x-y)$.
The path integral can then be formally expressed as:
\begin{equation}
\label{eq:part_func}
Z\[j;\mu\]= \[ \det 
\begin{pmatrix}
\B & \A \\
-\A & \B^{\dagger}
\end{pmatrix}
\]^{\frac{1}{2}}.
\end{equation}
We can rewrite the determinant as follows:
\begin{equation}\label{eq:det1}
\begin{split}
\det 
\begin{pmatrix}
\B & \A \\
-\A & \B^{\dagger}
\end{pmatrix}
&=
\det 
\begin{pmatrix}
-\A & \B^{\dagger} \\
\B & \A
\end{pmatrix}
=
 \exp {\rm Tr} \log
\begin{pmatrix}
-\A & \B^{\dagger} \\
\B & \A
\end{pmatrix}\\
&=  \exp {\rm Tr} \sum_{n} \dfrac{(-1)^{n+1}}{n} \begin{pmatrix}
-\A-\1 & \B^{\dagger} \\
\B & \A-\1
\end{pmatrix}^{n},
\end{split}
\end{equation}
where we used that the spin $1/2$ Dirac representation is even dimensional.
We arrive at the formula:
\begin{equation}
\log Z\[j;\mu\]= \dfrac{1}{2} {\rm Tr} \sum_{n} \dfrac{(-1)^{n+1}}{n} \begin{pmatrix}
-\A-\1 & \B^{\dagger} \\
\B & \A-\1
\end{pmatrix}^{n}.
\end{equation}

By induction on $n$ it is easy to show that the matrix 
$M_{n}\equiv\begin{pmatrix}
-\A- \1 & \B^{\dagger} \\
\B & \A- \1
\end{pmatrix}^{n}$ is of the form 
\begin{equation}
M_{n} = \begin{pmatrix}
p_{1}(\B^{\dagger} \B) & \B^{\dagger} \, p_{2}(\B^{\dagger} \B,\B \B^{\dagger})  \\
\B \, p_{3}(\B^{\dagger} \B,\B \B^{\dagger})  & p_{4}(\B \B^{\dagger}) 
\end{pmatrix},
\end{equation}
where $p_{i}$ are formal polynomials which can also depend on the differential operator $\A$. 
Therefore, evaluating the trace, the result is of the form
\begin{equation}\label{eq:expansion_j}
\log Z\[j;\mu\]= {\rm const} + \mathcal{O}\( j^{*} j\).
\end{equation}
This implies that the vacuum expectation value in equation~\eqref{vev} is zero:
\begin{equation}\label{eq:SSB_derivative}
\langle\delta \hat O_{\rm Maj}\rangle_{\mu} = \dfrac{\delta \log Z_{\mu}}{\delta j } \Bigg\vert_{j=0}=0.
\end{equation}
In this derivation we assumed that the expansion~\eqref{eq:expansion_j} is well defined, or in other words that the function is differentiable in $(j,j^*)=(0,0)$. For the free fermion we will explicitly evaluate the first two terms of this expansion in section~\eqref{sec:jj_term}, verifying the validity of this assumption.

The argument we presented relies on the fact that the Majorana source term has elements only on the principal diagonal of the quadratic form in equation~\eqref{quad}, whereas the kinetic term and the term describing the chemical potential are off-diagonal. It can be easily generalized to order parameters of the form $\delta O= c_{1} \overline{\Psi^{c}} \;\Psi + c_{2} \overline{\Psi^{c}} \,\gamma^{5}\,\Psi + \rm h.c.\,$.

\section{Free fermions at finite density}
\label{sec:fermionsfinitemu}

To gain better control of these formal manipulations we would like to explicitly compute the partition function of a system of free fermions at finite $\mu$ and zero temperature in the path integral approach. In the case of free bosons at finite $\mu$, the path integral with the $i \varepsilon$ prescription is convergent only for $\mu < m$~\cite{Nicolis:2023pye}. It is convenient to define $D_\mu$ to be a formal $\mu$-dependent covariant derivative. In the scalar case the computation reduces to the evaluation of the (inverse) determinant of the operator $-D_{\mu}D^{\mu} - m^2$, which can be easily seen to be independent of $\mu$ for $\mu < m$. On the other hand, for $\mu > m$ the gaussian path integral is divergent and one is forced to include interactions in order to stabilize the system for values of $\mu>m$. The situation is however different in the case of fermions: the quadratic fermionic path integral is always equal (by definition) to a functional determinant, and for a (constant and homogeneous) chemical potential one obtains again the operator $-D_{\mu}D^{\mu} - m^2$, both for $\mu < m$ and for $\mu > m$. This determinant should be independent of $\mu$ for $\mu < m$, essentially for the same reasons as in the scalar case, but should depend explicitly on $\mu$ in the regime $\mu > m$, corresponding to the free Fermi gas phase. How can these two conditions be mutually consistent? How does the $\mu$ dependence in the determinant arise? In order to answer these questions we shall carefully analyze the $i\varepsilon$ term that selects the finite density ground state, and compute the relativistic Fermi gas free energy in the path integral approach.

We start from the theory of a free Dirac fermion of mass $m$, as in eq.~\eqref{eq:Dirac_free}.
We are interested in studying this system at zero temperature, in the presence of a chemical potential for the vectorial charge (i.e., at finite charge density). To describe this system we first switch to the canonical formalism and include a term describing the system at finite chemical potential.
In canonical quantization the conjugate variables are
\begin{equation}
\Psi , \qquad \Pi = \dfrac{\partial \mathcal{L}}{\partial \dot{\Psi}}=\bar{\Psi} i \gamma^{0} = i \Psi^{\dagger}.
\end{equation}
The Legendre transform of the Lagrangian~\eqref{eq:Dirac_free} gives the canonical Hamiltonian density for a free Dirac fermion:
\begin{equation}
\begin{split}
\mathcal{H}=&\int \(\Pi \dot{\Psi}- \mathcal{L} \) = \int \(\Pi \dot{\Psi} - \Pi \dot{\Psi} - \bar{\Psi} i \gamma^{i}\partial_{i} \Psi + m \bar{\Psi} \Psi  \)\\
=&\int -\(\Pi \gamma^{0} \vec{\gamma}\cdot \vec{\nabla}\Psi + i m \Pi \gamma^{0} \Psi \) \\
=&\int \Pi \, \gamma^{0} \( - \vec{\gamma}\cdot \vec{\nabla}-i m \) \Psi.
\end{split}
\end{equation}
The system at finite chemical potential can then be described by the effective Hamiltonian
\begin{equation}
H_{\mu} = H- \mu Q,
\end{equation}
where $Q=\int \d^{3} x \;\bar{\Psi} \gamma^{0} \Psi$.
Going back to the Lagrangian formalism we obtain
\begin{equation}\label{eq:free}
\mathcal{L}_{\mu} = \bar{\Psi }\( i \slashed{\partial} + \gamma^{0} \mu - m \) \Psi .
\end{equation}

\subsection{$i \varepsilon$ term for the fermions}

To describe a finite density phase we want to project on the ground state of the operator~$H_\mu$. In order to do so we introduce an $i \varepsilon$ term in the Hamiltonian path integral for the fermions. We add to the action functional a term $i \varepsilon \mathcal{H}_{\mu}$, and obtain:
\begin{equation}
Z\[\mu\]  = \int D \Pi D \Psi \exp i \int_{t}^{t'} {\rm d}t \int {\rm d^{3}}x \left[ \Pi \dot{\Psi}- (1-i \varepsilon)
\mathcal{H}_{\mu}\right].
\end{equation}
This is equivalent to performing a Wick rotation $t \rightarrow (1-i\varepsilon)t$, as evidenced by rewriting the functional integral as
\begin{equation}
Z\[\mu\]  = \int D \Pi D \Psi \exp i \int_{t}^{t'} {\rm d}t (1-i \varepsilon) \int {\rm d^{3}}x \left[\dfrac{1}{ (1-i \varepsilon)} \Pi \dot{\Psi}-
\mathcal{H}_{\mu}\right].
\end{equation}
The quadratic path integral for a complex fermion is equal to the determinant of the kinetic operator. Up to an overall constant factor, we therefore obtain
\begin{equation}
Z\[\mu\]  = \det\left(  i \dfrac{\gamma^{0}\partial_{0}}{(1-i\varepsilon)} - i\gamma^{i}\partial_{i}+ \gamma^{0} \mu - m  \right),
\end{equation}
both for $\mu>m$ and for $\mu < m$.

Defining for convenience a covariant derivative $D_{\alpha} = \partial_{\alpha} - i \mu \, \delta_{\alpha 0}$, and working at linear order in $\varepsilon$, we arrive at
\begin{equation}
Z\[\mu\]  = \det\left(  i \slashed{D} - m - \varepsilon \gamma^{0} \partial_{0} \right).
\end{equation}

\subsection{Quantum mechanical fermionic oscillator}

As a warm-up, let us first consider the quantum mechanical fermionic oscillator, \emph{i.e.} QFT in $0+1$ dimensions.\footnote{In this quantum mechanical example we simply consider a \emph{spin $0$} fermionic particle, neglecting the (internal) spin degrees of freedom.} In this case there is no $\gamma$ matrix and we simply need to compute 
\begin{equation}
Z\[\mu\]  = \det\left( (1+ i \varepsilon) i \partial_{0} + \mu - m\right).
\end{equation}
Let us consider the charge density operator $\hat Q$. Taking a Fourier transform and working at linear order in $\varepsilon$ we have
\begin{align}
\langle \hat Q \rangle &= - i  \dfrac{\partial \log Z\[\mu\] }{\partial \mu} = - i \int \dfrac{{\rm d}\omega}{2\pi} \dfrac{1}{\omega +(\mu -m)(1-i \varepsilon)} \nonumber \\
&= - i \int \dfrac{{\rm d}\omega}{2\pi} \dfrac{1}{\omega +(\mu -m)-i \varepsilon \, {\rm sign}(\mu -m)}.
\end{align}
Using the distributional identity 
\begin{equation}
 \dfrac{1}{\omega +(\mu -m)-i \varepsilon(\mu -m)} = {\rm PV} \left( \dfrac{1}{\omega + \mu -m} \right) + i \pi \, {\rm sign}(\mu - m) \delta(\omega + \mu -m), 
\end{equation}
and integrating we obtain the charge density 
\begin{equation}
\langle \hat Q \rangle = -\dfrac{1}{2} + \theta(\mu -m).
\end{equation}
The $-1/2$ term is an additive constant that can be renormalized away. In quantum mechanics the charge operator is proportional to the Hamiltonian (but dimensionless), so that it corresponds to nothing else but the zero point energy of a fermionic quantum oscillator:
\begin{equation}
\hat{H} = \dfrac{1}{2} \hbar \omega [\hat{c}^{\dagger}, \hat{c}] = \hbar \omega \left( \hat{N}- \dfrac{1}{2} \right).
\end{equation}
The eigenvalue of the number operator $\hat{N}$ is $1$ for $\mu>m$, corresponding to the full ``Fermi sea'' of a nonzero charge state, and $0$ for $\mu<m$, corresponding to the zero charge ground state. 

\subsection{The Fermi gas in 3+1 dimensions}
\label{subsec:free}

In the case of a Dirac fermion in $3+1$ dimensions we need to deal with the spinor structure. Using charge conjugation and the transformation property of the $\gamma$ matrices $C \gamma_{\mu} C^{-1} = - (\gamma_{\mu})^{T}$, we have the identity
\begin{equation}
\begin{split}
Z\[\mu\] &=\det\left(  i \slashed{D} - m - \varepsilon \gamma^{0} \partial_{0} \right) = {\rm det}\left( C \left(  i \slashed{D} - m - \varepsilon \gamma^{0} \partial_{0}\right) C^{-1} \right)\\
&= \det\left( - i \slashed{D}^{T} - m + \varepsilon (\gamma^{0})^{T} \partial_{0} \right) = \det\left( - i \slashed{D} - m + \varepsilon \gamma^{0} \partial_{0} \right),
\end{split}
\end{equation}
where we used $(\det C)(\det C^{-1}) =1 $, and that the determinant is preserved by transposition.
Taking the geometric average, using the fact that the Dirac matrices are even dimensional, and neglecting terms of order $\varepsilon^{2}$ we obtain:
\begin{equation}
\begin{split}
Z\[\mu\] &=\det\left(\left(  i \slashed{D} - m - \varepsilon \gamma^{0} \partial_{0} \right) \left( i \slashed{D} + m - \varepsilon \gamma^{0} \partial_{0} \right)\right)^{1/2}\\
&= \det\left( -\slashed{D}\slashed{D} - m^{2} - i\varepsilon (\slashed{D} \gamma^{0} + \gamma^{0} \slashed{D})\partial_{0}\right)^{1/2} = \det\left( -\slashed{D}\slashed{D} - m^{2} -2 i\varepsilon D_{0}\partial_{0}\right)^{1/2}.
\end{split}
\end{equation}
For a constant and homogeneous chemical potential we can use the identity $\slashed{D}\slashed{D}= D_{\mu}D^{\mu}$. Taking a Fourier transform and using $(\log \det = \rm{Tr} \log)$, we arrive at
\begin{equation}\label{eq:integral_free}
\log Z\[\mu\] = 4\times \dfrac{1}{2} \int \dfrac{{\rm d}^4 p}{(2\pi)^{4}} \log \left( P^2 - m^{2} \right),
\end{equation}
where
\begin{equation}
P_\mu = \Big( (1+i \varepsilon) p_{0} + \mu, \,\vec{p}\, \Big) \qquad P^2 =\left[(1+i \varepsilon) p_{0} + \mu \right]^{2} - \vec{p}^{\,2},
\end{equation}
and the factor of $4$ comes from the trace in Dirac space. Notice that in relating the determinant of the Dirac operator to the determinant of an operator which is a multiple of the identity in Dirac space we have already made use of charge conjugation. The resulting singularity structure is asymmetric under $\mu\to -\mu$, but the conjugate contribution is already accounted for by the overall factor of $2$.

Let us analyze in more detail the singularities of the integrand in the complex $p_{0}$ plane, for fixed spatial momentum $\vec{p}$: there is a branch cut joining the two endpoints defined by the condition $P^2 -m^2= 0$,
\begin{equation}\label{eq:poles_epsilon_tilted}
p_{0,\pm} = \left(-\mu \pm \sqrt{\vec{p}^{\,2}+m^{2}}\right)(1-i\varepsilon).
\end{equation}
It is technically convenient to slightly deform the $i\varepsilon$ term in such a way that the singularities are all shifted by the same amount, with the direction of the shift determined by~\eqref{eq:poles_epsilon_tilted}. After this deformation, we have 
\begin{equation}\label{eq:poles_epsilon}
p_{0,\pm} = -\mu \pm \sqrt{\vec{p}^{\,2}+m^{2}} - i\varepsilon \, {\rm sign}\left(-\mu \pm \sqrt{\vec{p}^{\,2}+m^{2}}\right).
\end{equation}
It is straightforward to check that the integrand in eq.~\eqref{eq:integral_free} is single-valued once this branch cut is fixed.

For $\mu < m$ the branch cut endpoints $p_{0,\pm}$ lie in opposite quadrants (the second and the fourth). Let us choose the cut so that $p_{0,\pm}$ are joined through the point at infinity, without touching the first and third quadrants and the vertical strip comprised between $p_{0,\pm}$. With this choice we are free to deform the $p_{0}$ integration contour as follows: first Wick rotate from the real axis to the vertical $i p_{0}$ axis and then translate to the left by $\mu$. Doing this we do not encounter any singularity and it is easy to check that the contributions from the contours at infinity cancel each other. We are left with an integral on the vertical line, with two singularities lying at a distance $\pm \sqrt{\vec{p}^{\,2}+m^{2}}$, on the left and the right. This can be recognized as the Euclidean path integral of the $\mu=0$ theory. Therefore we have 
\begin{equation}
f(\mu)=0 \qquad {\rm for} \qquad \mu < m.
\end{equation}
This is reminiscent of the so-called ``Silver Blaze'' property of QCD at (small) nonzero isospin chemical potential~\cite{Cohen:2003kd}. Notice, however, that our chemical potential is associated to the $\U(1)$ vectorial symmetry and is thus more similar to a baryon chemical potential.

\begin{figure}
\begin{center}
\begin{overpic}[width=.49\textwidth]{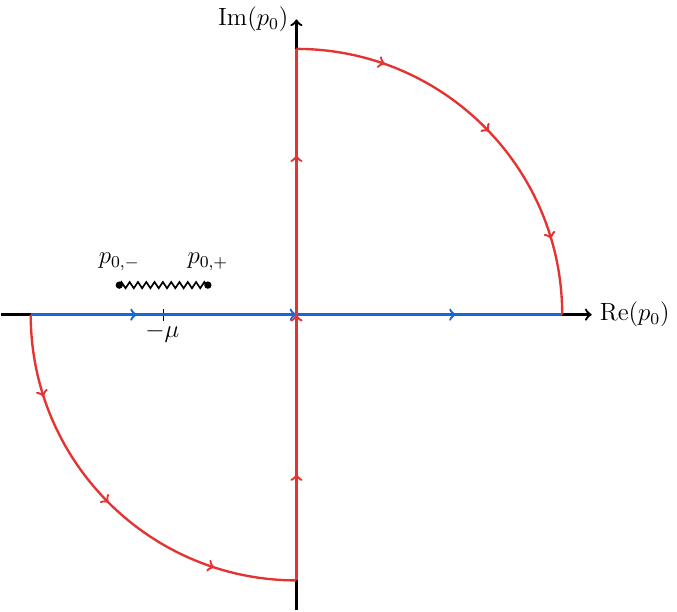}
\put(90,-15){a)}
\end{overpic}
\begin{overpic}[width=.49\textwidth]{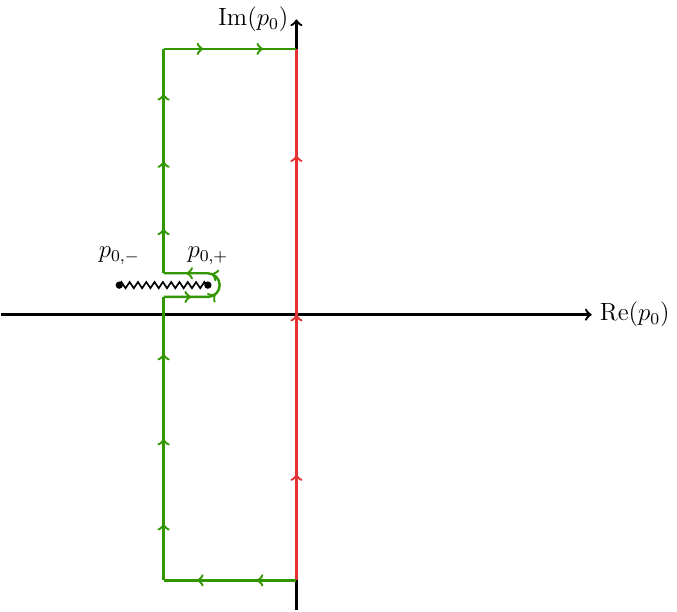}
\put(90,-15){b)}
\end{overpic}
\end{center}
\caption{\label{fig:contour} \it \small Analytic structure and contour deformation for $\mu>m$ and $|\vec{p}| < p_{F}$. In the first step, we deform the integration contour from the real axis (blue) to the imaginary axis (red), see panel~a). The deformation is smooth and the integrals along the two arcs in the first and third quadrants cancel each other. In the second step, we deform the contour with a shift by $-\mu$. The integral along the vertical green line reproduces the $\mu=0$ path integral, while the contributions from the discontinuity across the branch cut and the segments at infinity reproduce the finite $\mu$ free energy density of the relativistic Fermi gas, see eq.~\eqref{eq:ffree}.}
\end{figure}

For $\mu > m$, the location of the endpoints depends on the relative size of $|\vec{p}|$ and the Fermi momentum $p_{F}$:
\begin{equation}
p_{F}\equiv \sqrt{\mu^{2}-m^{2}} .
\end{equation}
They lie in opposite quadrants (the second and the fourth) only when $|\vec{p}| > p_{F}$. In this case the previous analysis is unchanged. For $|\vec{p}| < p_{F}$, instead, both endpoints lie in the second quadrant and more care is needed. In this case we choose the cut to be the horizontal segment joining the two points, see fig.~\ref{fig:contour}. Trying to follow the same contour deformation procedure as before we obtain some additional contributions due to the singularities. The Wick rotation goes smoothly and one can easily check that also in this case the contributions from the large arcs cancel each other (left panel of fig.~\ref{fig:contour}). The translation to the left by $\mu$ gives instead two contributions: one ($I_1$) from the discontinuity along the branch cut of the log; the other ($I_2$) from the integrals along the segments $({\rm Re}(p_{0}),{\rm Im} (p_{0})) = ([-\mu,0],\pm i \Lambda)$, see the right panel of fig.~\ref{fig:contour}. We have:
\begin{equation}
\log \dfrac{Z\[\mu\] }{Z\[0\] } = 2  \int_{|\vec{p}|<p_{F}} \dfrac{{\rm d}^{3}\vec{p}}{(2\pi)^{3}} \left( I_1+  I_2 \right),
\end{equation}
where
\begin{equation}
\begin{split}
&I_1 = \int_{-\mu}^{-\mu + \sqrt{\vec{p}^{\, 2}+m^{2}}} \dfrac{{\rm d}p_{0}}{2\pi} (-2 \pi i) = - i \sqrt{\vec{p}^{\, 2}+m^{2}},\\
&I_2 = \lim_{\Lambda \rightarrow \infty} \left[ \int_{0- i \Lambda}^{-\mu - i \Lambda} \dfrac{{\rm d}p_{0}}{2\pi} \log\left(e^{-i\pi} \Lambda^{2} \right) + \int_{-\mu+ i \Lambda}^{0+ i \Lambda} \dfrac{{\rm d}p_{0}}{2\pi} \log\left(e^{i\pi} \Lambda^{2} \right) + \mathcal{O}\left(\dfrac{1}{\Lambda }\right) \right]= + i \mu .
\end{split}
\end{equation}
In the second equation we paid particular attention to the definition of the phase with respect to the cut.
We thus arrive at:
\begin{equation}
\label{eq:Fermi_gas}
f(\mu) = 2 \int_{|\vec{p}|<p_{F}} \dfrac{{\rm d}^{3}\vec{p}}{(2\pi)^{3}} \left( \sqrt{\vec{p}^{\, 2}+m^{2}} - \mu \right)  \qquad {\rm for} \qquad \mu > m,
\end{equation}
which can be recognized as the zero temperature limit of the free energy density of a relativistic Fermi gas.
The integral can be computed explicitly and we obtain:
\begin{equation}\label{eq:ffree}
f(\mu) =  \Theta\Big(\mu^2-m^2\Big) \dfrac{1}{24\pi^{2}}\left(  \mu (5m^{2}-2\mu^{2})\sqrt{\mu^{2}-m^{2}} - 3 m^{4} \arctanh\left(\dfrac{\sqrt{\mu^{2}-m^{2}}}{\mu}\right) \right).
\end{equation}
The charge density is simply given by
\begin{equation}
\langle \hat J^{0} \rangle = - \dfrac{\partial f(\mu)}{\partial \mu} = \dfrac{p_{F}^{3}}{3\pi^{2}}, \qquad p_{F}= \sqrt{\mu^{2}-m^{2}}.
\end{equation}
The same result for the charge density can be obtained by first performing the derivative and then carrying out the contour integration (in which case only a pole is present, instead of a branch cut). These equations match the well-known results for the relativistic Fermi gas, usually derived in the canonical formalism.

\subsection{Absence of symmetry breaking for free fermions}
\label{sec:jj_term}

In this section we try to compute explicitly the partition function in presence of a Majorana source term, using equations~\eqref{eq:part_func} and~\eqref{eq:det1} as a starting point:
\begin{equation}
Z\[j;\mu\] = \[ \det 
\begin{pmatrix}
\B & \A \\
-\A & \B^{\dagger}
\end{pmatrix}
\]^{\frac{1}{2}}= \[ \det 
\begin{pmatrix}
-\A & \B^{\dagger} \\
\B & \A
\end{pmatrix}
\]^{\frac{1}{2}},
\end{equation}
where
\begin{equation}
\A = [ -i \slashed{D}_{x} + m ]_{\alpha\beta} \, \delta^{4}(x-y), \qquad \B = j \, [\gamma^{2}\gamma^{0}]_{\alpha\beta} \, \delta^{4}(x-y),
\end{equation} 
are $4\times 4$ matrix operators.\footnote{Here we have used $(\gamma^{2})^{*}=-\gamma^{2}$, valid in the chiral basis discussed in appendix~\ref{app:conventions}.}

The operator $\A$ is invertible, therefore we can consider the decomposition:
\begin{equation}
\begin{pmatrix}
-\A & \B^{\dagger}\\
\B & \A 
\end{pmatrix} =
\begin{pmatrix}
-\A & \0 \\
\B & \1
\end{pmatrix}
\begin{pmatrix}
\1 & - \A^{-1} \B^{\dagger} \\
\0 & \A +\B \A^{-1} \B^{\dagger}
\end{pmatrix},
\end{equation}
from which it follows that
\begin{equation}
\begin{split}
\det 
\begin{pmatrix}
-\A & \B^{\dagger}\\
\B & \A 
\end{pmatrix}&=\det(-\A) \det(\A +\B \A^{-1} \B^{\dagger})
= \det(\A^2 +\A \B \A^{-1} \B^{\dagger})\\
&= \det(\A^2)\det(\1 +\A^{-1} \B \A^{-1} \B^{\dagger}).
\end{split}
\end{equation}
The first factor corresponds to the usual $j=0$ fermionic determinant (see e.g.~the previous section), and we can thus focus on the second term that will give rise to the finite $j$ corrections. Since the operator $\A$ is independent of $j,j^*$ and the operator $\B$ is proportional to $j$, we see that the partition function can depend on $j$ and $j^*$ only through the combination $j j^{*}$, in agreement with the argument of section~\ref{sec:SSB}.

The operators $\B,\B^{\dagger}$ can be rewritten as 
\begin{equation*}
\B =  j\, \left(-i \, C\right) \, \delta^{4}(x-y) , \qquad \B^\dagger = j^*\, \left(i \, C^{\dagger}\right) \, \delta^{4}(x-y),
\end{equation*}
allowing to express the second determinant as 
\begin{equation}
\det(\1 +\A^{-1} \B \A^{-1} \B^{\dagger})= \det(\1 + jj^*\A^{-1} (\A^{c})^{-1} ).
\end{equation}
In order to show that no subtlelties arise in equation~\eqref{eq:SSB_derivative} we want to show that the first order term in the $jj^*$ expansion of the logarithm of the partition function, $\log Z_{\mu}\[j,j^{*}\]$, is regular up to the usual UV renormalization. In other words, defining
\begin{equation}
\delta_{jj^{*}}[\mu]\equiv\dfrac{\rm d}{{\rm d}(j j*)}\log Z\[j;\mu\]  \Bigg\vert_{j=0}= \dfrac{1}{2} \dfrac{\rm d}{{\rm d}(j j*)} \log \det(\1 + jj^*\A^{-1} (\A^{c})^{-1} ) \Bigg\vert_{j=0}
\end{equation}
we would like to show that $\delta_{jj^{*}}[\mu]- \delta_{jj^{*}}[0]$ is finite. After straightforward manipulation we obtain
\begin{equation}
\delta_{jj^{*}}[\mu] = \dfrac{1}{2} \Tr\left( \A^{-1} (\A^{c})^{-1} \right)=  \dfrac{1}{2} \int \dfrac{{\rm d}^4 p}{(2\pi)^{4}} {\rm tr} \left(-\dfrac{1}{(\slashed{P})^T + m}\right) \left(\dfrac{1}{\slashed{P} - m} \right).
\end{equation}
Carrying out the trace in Dirac space and using the transpose of the $\gamma$ matrices (see appendix~\ref{app:conventions}) we arrive at
\begin{equation}
\delta_{jj^{*}}[\mu] = - \dfrac{1}{2} \int \dfrac{{\rm d}^4 p}{(2\pi)^{4}} {\rm tr} \left(\dfrac{(\slashed{P})^T - m}{P^2-m^2}\right) \left(\dfrac{\slashed{P} +m}{P^2-m^2} \right) =  - 4\times \dfrac{1}{2} \int \dfrac{{\rm d}^4 p}{(2\pi)^{4}} \left(\dfrac{P^2 - m^2 + 2 p_2^2}{(P^2-m^2)^2}\right) .
\end{equation}
We can now compute $\delta_{jj^{*}}[\mu]- \delta_{jj^{*}}[0]$ using the contour integration approach previously described, obtaining
\begin{equation}
\delta_{jj^{*}}[\mu] - \delta_{jj^{*}}[0] =  + 2i\, \int_{|\vec{p}|<p_{F}} \dfrac{{\rm d}^{3}\vec{p}}{(2\pi)^{3}} \; {\rm Res}_{\,p_{0,+}} \left(\dfrac{1}{P^2-m^2}+\dfrac{2 p_2^2}{(P^2-m^2)^2}\right) .
\end{equation}
The second term in parenthesis has a double pole in $p_{0,+}$ and has therefore vanishing residue. 
Evaluating the residue of the first term, we arrive at
\begin{equation}
\begin{split}
\delta_{jj^{*}}[\mu] - \delta_{jj^{*}}[0] &=  i\,  \int_{|\vec{p}|<p_{F}} \dfrac{{\rm d}^{3}\vec{p}}{(2\pi)^{3}} \dfrac{1}{\sqrt{\vec{p}^{\, 2}+m^{2}}} \\
&= i\,  \, \dfrac{1}{4\pi} \left(p_{F} \sqrt{p_{F}^{2}+m^{2}} - m^2 \, \arctanh\left(\dfrac{p_F}{\sqrt{p_{F}^{2}+m^{2}}}\right) \right),
\end{split}
\end{equation}
which is manifestly finite and regular.

We can thus conclude that 
\begin{equation}
\log Z\[j;\mu\] - \log Z\[j;0\] = - i  \, f(\mu) + \Big( \delta_{jj^{*}}[\mu] - \delta_{jj^{*}}[0] \Big) j j^* + \mathcal{O}\Big( (j j^*)^2 \Big),
\end{equation}
where the first two coefficients are finite. We can thus conclude that the finite density phase does not induce an expectation value for the Majorana operator, and therefore that the ${\rm U(1)_{V}}$ global symmetry is unbroken. This explicit computation puts on firmer grounds the naive argument of section~\ref{sec:SSB} in the case of free fermions.

\section{The Fermi gas in a magnetic background}
\label{sec:magnetic}

Having understood how to deal with free fermions at finite density in the path integral approach, we would like now to consider the dynamics of electrons in a magnetic background, \emph{i.e.} finite density QED with an external magnetic field.

\subsection{The functional determinant}

Turning on gauge interactions, we consider the fermionic path integral in a magnetic background.
The covariant derivative is now given by 
\begin{equation}
D_{\alpha} = \partial_{\alpha} - i e A_{\alpha} - i \mu \delta_{\alpha 0},
\end{equation}
where $e$ is the QED coupling constant and $A$ is a background gauge field associated to a homogeneous and constant magnetic background.
Following the same steps as in the previous case, the fermion one-loop contribution to the partition function is given by:
\begin{equation}
\label{eq:Z}
Z^{(1)}[A;\mu]= \det\left( -\slashed{D}\slashed{D} - m^{2} -2 i\varepsilon D_{0}\partial_{0}\right)^{1/2},
\end{equation}
where now $\slashed{D} \slashed{D} = D_{\mu}D^{\mu}- \frac{e}{2} \sigma_{\mu\nu} F^{\mu\nu}$. 
The differential operator $D_{\mu}D^{\mu}$ acts on a test function $f(x)$ as:
\begin{equation}
D_{\mu}D^{\mu} f = \Box f -2 i e A_{\mu} \partial^{\mu}f -4 \mu e A^{0} f -2i \mu \partial^{0} f - e^{2} A_{\mu}^{2} f - \mu^{2} f - i e (\partial_{\mu} A^{\mu}) f.
\end{equation}
For a homogeneous and constant external field we can choose a gauge in which $A_{0}=0$ and $\partial_{\mu}A^{\mu}=0$, so that this expression is simplified. 
Let us consider the case of a magnetic field $\vec{B}= B\hat{z}$ along the $z$-axis and choose $\vec{A} =B x \,\hat{y}$. Using the explicit expression and reorganizing the terms, the partition function~\eqref{eq:Z} is given by the determinant of the differential operator
\begin{equation}
L \equiv \left( i (1+i\varepsilon)\partial_{0}-\mu \right)^{2} - (i \partial_{x})^{2} - (i \partial_{z})^{2} - (i \partial_{y}+ eB x)^{2} -m^{2} + \dfrac{e}{2} \sigma_{\mu\nu} F^{\mu\nu}.
\end{equation}
In order to find its eigenvalues we use the method of separation of variables and look for eigenfunctions of the form
\begin{equation}
f(x,y,z,t)= e^{-i p_{0} t} e^{i p_{y} y} e^{i p_{z} z} g(x).
\end{equation}
This is an eigenfunction of $L$ provided that $g(x)$ satisfies:
\begin{equation}
\left(\partial_{x}^{2}-(p_{y}+e B x )^{2}\right) g(x) = \lambda_{g} g(x).
\end{equation}
Changing variable to $\xi =\sqrt{eB} x + \dfrac{1}{\sqrt{eB}} p_{y}$, the equation becomes 
\begin{equation}
e B \left(\partial_{\xi}^{2}- \xi^{2}\right) g(\xi)= \lambda_{g} g(\xi),
\end{equation}
with regular solutions
\begin{equation}
\begin{split}
&g_{n}(\xi)= H_{n}(\xi) e^{-\xi^{2}/2},\\
&\lambda_{n} = -eB(2n+1),
\end{split}
\end{equation}
where $H_{n}(\xi)$ are the Hermite polynomials. We assumed implicitly $eB>0$ for notational simplicity, otherwise $eB$ should be replaced by $\vert eB \vert$. 
We have thus found the eigenvalues of $L$. Denoting the spin with $\sigma=\pm 1/2$:
\begin{equation}
\lambda^{(L)}_{n,p_{0},p_{y},p_{z}}= \left((1+i\varepsilon)p_{0}+\mu \right)^{2} - (p_{z})^{2} - eB(2n+1) -m^{2} + 2 \sigma e B.
\end{equation}
Taking into account the multiplicities and the normalization of the eigenfunctions $g_{n}$, the partition function is given by
\begin{equation}
\label{eq:Zmu}
\begin{split}
\log Z^{(1)}[A;\mu]= &\dfrac{e B }{2\pi} \int \dfrac{{\rm d}p_{0}}{2\pi} \dfrac{{\rm d}p_{z}}{2\pi}  \\&\sum_{\sigma = \pm \frac{1}{2}} \sum_{n=0}^{\infty}  \log \left( 
\left((1+i\varepsilon)p_{0}+\mu \right)^{2} - (p_{z})^{2} - eB(2n+1) -m^{2} + 2 \sigma e B
\right).
\end{split}
\end{equation}

\subsection{A modern derivation of the nonperturbative Euler--Heisenberg effective action}

Before discussing the finite $\mu$ case, let us first rederive the one-loop QED effective action in a background homogeneous magnetic field. 
We start from the expression of the partition function~\eqref{eq:Zmu}, evaluated at $\mu=0$. In this case we can safely perform a Wick rotation to Euclidean momenta and use the identity $\log(p^{2}+k)= -\partial_\alpha(p^{2}+k)^{-\alpha}\big\vert_{\alpha=0}$. The resulting expression involves an integral in ${\rm d^{2}}p$ which we evaluate in dimensional regularization to obtain
\begin{equation}
\begin{split}
\log Z^{(1)}[A;0] 
&=- \dfrac{\partial}{\partial \alpha}\Bigg\vert_{0} \dfrac{e B }{2\pi}\sum_{\sigma = \pm \frac{1}{2}} \sum_{n=0}^{\infty} \int \dfrac{{\rm d^{d}}p}{(2\pi)^{d}}   \left(\dfrac{1}{ p^{2} - eB(2n+1-2\sigma) -m^{2}}\right)^{\alpha} \\
&=- i\dfrac{e B }{2\pi} \dfrac{1}{(4\pi)^{d/2}} \sum_{\sigma = \pm \frac{1}{2}}\dfrac{\partial}{\partial \alpha}\Bigg\vert_{0} \left( \dfrac{\Gamma(\alpha-\frac{d}{2})}{\Gamma(\alpha)} \sum_{n=0}^{\infty}  \dfrac{1}{ (2eB)2n+(1-2\sigma)eB +m^{2}} \right) \\
&=- i\dfrac{e B }{2\pi} \dfrac{1}{(4\pi)^{d/2}} \sum_{\sigma = \pm \frac{1}{2}}\dfrac{\partial}{\partial \alpha}\Bigg\vert_{0} \left( \dfrac{\Gamma(\alpha-\frac{d}{2})}{\Gamma(\alpha)} (2 e B)^{-\alpha+\frac{d}{2}} \zeta\left(\alpha-\dfrac{d}{2},\dfrac{m^{2}}{2eB}+\dfrac{1}{2}-\sigma\right) \right),
\end{split}
\end{equation}
where $\zeta(s,q)$ is the Hurwitz zeta function defined by 
\begin{equation}
\zeta(s,q)=\sum_{n=0}^{\infty} \dfrac{1}{(n+q)^{s}}.
\end{equation}
Evaluating the derivative at $\alpha=0$, we arrive at
\begin{equation}
\log Z^{(1)}[A;0]= - i\dfrac{e B }{2\pi} \left(\dfrac{2 e B}{4\pi}\right)^{\frac{d}{2}} \Gamma\left(-\dfrac{d}{2}\right)  \left[\zeta\left(-\dfrac{d}{2},\dfrac{m^{2}}{2eB}\right)+\zeta\left(-\dfrac{d}{2},\dfrac{m^{2}}{2eB}+1\right)\right].
\end{equation}
From the definition of the Hurwitz zeta function we have the identity:
\begin{equation}
\zeta(s,q+1) = \zeta(s,q) - \dfrac{1}{q^{s}},
\end{equation}
which lets us rewrite the effective action as
\begin{equation}
\log Z^{(1)}[A;0]= - i \dfrac{e B}{2\pi} \left(\dfrac{2 e B}{4\pi}\right)^{\frac{d}{2}} \Gamma\left(-\dfrac{d}{2}\right)  \left[2\zeta\left(-\dfrac{d}{2},\dfrac{m^{2}}{2eB}\right)-\left(\dfrac{m^{2}}{2eB}\right)^{\frac{d}{2}}\right].
\end{equation}
For $d=2-2\varepsilon$, in the limit $\varepsilon \rightarrow 0$:
\begin{equation}
\begin{split}
\log Z^{(1)}[A;0]=&\, i \dfrac{e^{2}B^{2}}{4\pi^{2}}\left(\dfrac{1}{\varepsilon}-\gamma_{E}+\log\left(4\pi\right)\right)\left[2\, \zeta\left(-1,\dfrac{m^{2}}{2eB}\right)-\dfrac{m^{2}}{2eB} \right]+ \\
&\hspace{-50pt}+ i  \dfrac{e^{2}B^{2}}{4\pi^{2}} \left[2 \, \zeta' \left(-1,\dfrac{m^{2}}{2eB}\right)+\left(1- \log\left(\dfrac{2eB}{\bar{\mu}^2}\right)\right)2 \, \zeta\left(-1,\dfrac{m^{2}}{2eB}\right)- \dfrac{m^{2}}{2eB} \left(1-\log\left(\dfrac{m^{2}}{\bar{\mu}^{2}}\right) \right)\right],
\end{split}
\end{equation}
where we denoted the (arbitrary) renormalization scale as $\bar{\mu}$, not to be confused with the chemical potential $\mu$, and $\zeta'(s,q)={\rm d}\zeta(s,q)/{\rm d}s$.
Using the identity
\begin{equation}
\label{eq:Bernoulli}
\zeta\left(-n,x\right)=-\dfrac{B_{n+1}(x)}{n+1} \Longrightarrow \zeta\left(-1,x\right)=-\dfrac{B_{2}(x)}{2}= -\dfrac{1}{12}+\dfrac{x}{2}-\dfrac{x^{2}}{2},
\end{equation}
where $B_{n}(x)$ are Bernoulli polynomials, the first (divergent) term can be rewritten as
\begin{equation}
\label{eq:div_EH}
i \left(\dfrac{1}{\varepsilon}-\gamma_{E}+\log\left(4\pi\right)\right) \left(-\dfrac{m^{4}}{16\pi^{2}}-\dfrac{e^{2}B^{2}}{24\pi^{2}} \right),
\end{equation}
showing explicitly that all the divergencies can be removed by a cosmological constant counter-term and a wave-function renormalization for the magnetic field $B$ or, equivalently, the field strength $F_{\mu\nu}$, which for vanishing electric field satisfies the relation $B^{2}=-F_{\mu\nu}F^{\mu\nu}/2$. We choose to work in the $\overline{\rm MS}$ renormalization scheme and remove all the terms in~\eqref{eq:div_EH}. For simplicity we also chose the renormalization scale to be $\bar{\mu}=m$.
We have obtained the full partition function in closed form, non-perturbatively in the coupling $e$ and in the non-dynamical external magnetic field $B$. It is convenient to express the result in terms of $\beta= 2eB/m^{2}$:
\begin{equation}
\label{eqref:eff_zeta}
\begin{split}
\log Z^{(1)}[A;0]=i \dfrac{m^{4}}{16\pi^{2}}\beta^{2} 
\Bigg[&2 \, \zeta' \left(-1,\dfrac{1}{\beta}\right)
+2 \Big(1-\log\beta\Big) \zeta\left(-1,\dfrac{1}{\beta}\right)- \dfrac{1}{\beta}
\Bigg],
\end{split}
\end{equation}
to be compared with the well-known result of Euler and Heisenberg~\cite{Heisenberg:1936nmg,Weisskopf:1936hya,Schwinger:1951nm}
\begin{equation}
\log Z^{(1)}[A;0]= \dfrac{i}{8\pi^{2}} \int_{0}^{\infty}\dfrac{{\rm d}s}{s^{2}} \left[eB \cot(eB s) -\dfrac{1}{s} +\dfrac{1}{3} e^{2}B^{2} s\right] e^{-i m^{2} s}.
\end{equation}
An equivalent result has been obtained long ago through zeta function regularization methods, see for instance chapter 6 of Ref.~\cite{Dittrich:1985yb}, and Refs.~\cite{Blau:1988iz,Dunne:2004nc} for a detailed discussion.
We can check explicitly that the zeta function representation agrees with the Euler--Heisenberg representation in the weak field limit, where $\beta= 2eB/m^{2}\rightarrow 0$. Using the asymptotic expansion (see section 25.11 of~\cite{NIST:DLMF}):
\begin{equation}
\zeta' \left(-1,\dfrac{1}{\beta} \right) \xrightarrow[\beta \rightarrow 0]{}\frac{1}{12} -\frac{1}{4\beta^{2}}-\left(\frac{1}{2\beta^{2}}-\frac{1}{2\beta}+\frac{1}{12}\right) \log\beta -\sum _{k=1}^{\infty} \frac{ B_{2 k+2}(0)}{(2 k+2) (2 k+1)2 k}\beta^{2 k},
\end{equation}
together with equation~\eqref{eq:Bernoulli}, we recover the expansion
\begin{equation}
\label{eq:weak_mag}
\begin{split}
-i\log Z^{(1)}[A;0]\Bigg\vert_{\beta\rightarrow 0}
&= -3 \dfrac{m^{4} }{32 \pi^{2}} - \dfrac{m^{4} }{8\pi^{2}} \sum _{k=1}^{\infty} \frac{ B_{2 k+2}(0)}{(2 k+2) (2 k+1)2 k}\beta^{2 k+2}\\
&=-3\frac{m^4}{32 \pi ^2} +\frac{\alpha ^2}{90 m^4} \mathcal{F}^4-\frac{4 \pi  \alpha ^3 }{315 m^8}\mathcal{F}^6+\frac{16 \pi ^2 \alpha ^4 }{315 m^{12}} \mathcal{F}^8+ \mathcal{O}(\alpha^{5}),
\end{split}
\end{equation}
where $\mathcal{F}^2=F_{\mu\nu}F^{\mu\nu}=-2B^{2}$ and $\alpha= e^{2}/4\pi$, which reproduces the $F_{\mu\nu}F^{\mu\nu}$ terms of the Euler--Heisenberg Lagrangian with the correct numerical coefficients. 

The zeta function representation we derived in equation~\eqref{eqref:eff_zeta}, allows to easily obtain the strong field limit $\beta= 2eB/m^{2}\rightarrow \infty$:
\begin{equation}\label{eq:strong_mag}
-i\log Z^{(1)}[A;0]\Bigg\vert_{\beta\rightarrow \infty} = \dfrac{e^{2}B^{2}}{24\pi^{2}} \left(\log\left(\dfrac{2eB}{m^{2}}\right)-1+12 \,\zeta'(-1) \right),
\end{equation}
where we chose again $\bar{\mu}= m$.
We see that in the strong field limit the only dependence on the fermion mass appears in the logarithmic running term, where $m$ plays the role of an IR regulator.

The purely electric case is obtained from the substitution $B\rightarrow i E$. The (non-perturbative) Schwinger effect~\cite{Schwinger:1951nm} is encoded in the imaginary part of the full result~\eqref{eqref:eff_zeta} and is not captured by the asymptotic expansion, which is real.

Adding the tree-level contribution, the free energy for a constant and homogeneous magnetic field minimally coupled to a charged Dirac fermion is given by 
\begin{equation}
f(B,0)= -\dfrac{B^2}{2} + i \log Z^{(1)}[A;0],
\end{equation}
which (for $\bar\mu=m$) is
\begin{equation}
\label{eq:fB}
f(B,0)= -\dfrac{B^2}{2} -\dfrac{m^{4}}{16\pi^{2}}\beta^{2} 
\Bigg[2 \, \zeta' \left(-1,\dfrac{1}{\beta}\right)
+2 \Big(1-\log\beta\Big) \zeta\left(-1,\dfrac{1}{\beta}\right)- \dfrac{1}{\beta}\Bigg],
\end{equation}
with $\beta= 2eB/m^{2}$.

\subsection{Finite density QED in a magnetic background}

Reintroducing the chemical potential we consider equation~\eqref{eq:Zmu}:
\begin{align}
\log Z^{(1)}[A;\mu]= &\,\dfrac{e B }{2\pi} \int \dfrac{{\rm d}p_{0}}{2\pi} \dfrac{{\rm d}p_{z}}{2\pi}  \\&\sum_{\sigma = \pm \frac{1}{2}} \sum_{n=0}^{\infty}  \log \left( 
\left((1+i\varepsilon)p_{0}+\mu \right)^{2} - (p_{z})^{2} - eB(2n+1) -m^{2} + 2 \sigma e B
\right). \nonumber
\end{align}
The $\log$ in the integral has branching points at 
\begin{equation}
p_{0,\pm} = \left(-\mu \pm \sqrt{p_{z}^{2}+e B (2n+1-2\sigma)+m^{2}}\right)(1-i\varepsilon).
\end{equation}
The two points $p_{0,\pm}$ lie in the same (second) quadrant when
\begin{equation}
\begin{split}
&n \leq n_{\rm max}=
\left\lfloor\dfrac{\mu^{2}-m^{2}+ eB (2\sigma -1)}{2eB}\right\rfloor,\\
&p_{z}^{2} \leq p_{\rm max}^{2}= \mu^{2}-m^{2}- e B (2n+1-2\sigma).
\end{split}
\end{equation}
We can now perform the integral by deforming the contour as detailed in the free Fermi gas and write the partition function as the sum of a term given by the partition function at $\mu=0$ and a term that we know analytically:
\begin{equation}
\begin{split}
\log Z^{(1)}[A;\mu] =& \log Z^{(1)}[A;0] \,+\\
&-i \dfrac{e B }{2\pi} \sum_{\sigma = \pm \frac{1}{2}} \sum_{n=0}^{n_{\rm max}}  \int_{0}^{p_{\rm max}} \dfrac{{\rm d}p_{z}}{2\pi}  \left( 
\sqrt{ (p_{z})^{2}- p_{\rm max}^{2} + \mu^{2} } - \mu
\right).
\end{split}
\end{equation}
Shifting the index $n$ by a unit for $\sigma=-\frac{1}{2}$ we can rewrite the result as 
\begin{equation}
\label{eq:integral}
\begin{split}
\log Z^{(1)}[A;\mu] =& \log Z^{(1)}[A;0] \,- i \dfrac{e B }{4\pi^{2}} \int_{0}^{\sqrt{\mu^{2}-m^{2}}} {\rm d}p_{z}  \left( 
\sqrt{ (p_{z})^{2} + m^{2}} - \mu
\right) \\
& - i \dfrac{e B }{2\pi^{2}} \sum_{\ell =1}^{\ell_{\rm max}}  \int_{0}^{\sqrt{\mu^{2}-m^{2}-2eB \ell}} {\rm d}p_{z}  \left( 
\sqrt{ (p_{z})^{2}+m^{2}+2eB \ell} - \mu
\right),
\end{split}
\end{equation}
where
\begin{equation}
\ell_{\rm max}=
\left\lfloor \dfrac{\mu^{2}-m^{2}}{2eB} \right\rfloor,
\end{equation}
and the sum over $\ell$ gives a sum over Landau levels. It is convenient to express the result in terms of the free energy density
\begin{equation}
f(B,\mu)=-\dfrac{B^2}{2}  + i \log Z^{(1)}[A;\mu].
\end{equation}
The integrals can be computed in closed form
\begin{equation}
\label{eq:full_bmu}
f(B,\mu)\Big\vert_{\rm full} = \,f(B,0) \,+ \dfrac{1}{2} F_0 
+ \sum_{\ell =1}^{\ell_{\rm max}} F_\ell \; ,
\end{equation}
where we defined
\begin{equation}
F_\ell \equiv \dfrac{e B}{4\pi^{2}} \left((m^{2}+2eB\ell)\,\arctanh\left(\dfrac{\sqrt{\mu^{2}-(m^{2}+2eB\ell)}}{\mu}\right)- \mu \sqrt{\mu^{2}-(m^{2}+2eB\ell)}\right).
\end{equation}
It is possible to rewrite this result in terms of logarithms using the identity 
\begin{equation}
\arctanh\left(\dfrac{\sqrt{\mu^{2}-m^{2}}}{\mu}\right) =\log\left(\dfrac{\mu+\sqrt{\mu^{2}-m^{2}}}{m}\right).
\end{equation}
In the strong field case, when $2eB > (\mu^{2}-m^{2})$ and $l_{\max}=0$, only the first term contributes and the full result takes a simple form
\begin{equation}
\begin{split}
f(B,\mu)\Big\vert_{\rm strong} =& \,-\dfrac{B^2}{2} + \dfrac{e^{2}B^{2}}{24\pi^{2}} \left(\log\left(\dfrac{2eB}{\bar\mu^{2}}\right)-1+12 \,\zeta'(-1) \right) \\
&+\dfrac{e B}{8\pi^{2}} \left(m^{2}\,\arctanh\left(\dfrac{\sqrt{\mu^{2}-m^{2}}}{\mu}\right)- \mu \sqrt{\mu^{2}-m^{2}}\right),
\end{split}
\end{equation}
where in the first line we assumed also $2eB \gg m^2$.
On the other hand, in the weak field limit $2eB \ll (\mu^{2}-m^{2})$ the sum can be recognized as a Riemann sum in the variable 
\begin{equation}
q^{2}_{\ell} = 2eB \ell, \qquad \Delta q^{2}_{\ell} = 2eB.
\end{equation}
It is convenient to add and subtract the $\ell =0$ term, so that in the limit $2eB \rightarrow 0$ the sum from $\ell =0$ to $\ell_{\rm max}$ converges to the definite integral in ${\rm d}q^{2}$ with extrema
\begin{equation}
q^{2}_{\rm min} =0, \qquad q^{2}_{\rm max} =2eB \Bigg\lfloor \dfrac{\mu^{2}-m^{2}}{2eB}\Bigg\rfloor \xrightarrow[2 e B \rightarrow 0]{} (\mu^{2}-m^{2}).
\end{equation}
The leading order result becomes more transparent by taking a step back and writing the terms of the sum as integrals in ${\rm d}p_{z}$, as in equation~\eqref{eq:integral},
\begin{equation}
\begin{split}
f(0,\mu)= \dfrac{1}{4\pi^{2}} \int_{0}^{\mu^{2}-m^{2}} {\rm d}q^{2} \int_{0}^{\sqrt{\mu^{2}-m^{2}-q^{2}}} {\rm d}p_{z}  \left( 
\sqrt{ (p_{z})^{2}+q^{2}+m^{2}} - \mu \right) + {\rm const}+ \mathcal{O}\left(e B\right).
\end{split}
\end{equation}
One can now recognize the integral as an integral in cylindrical coordinates over a domain which is a ball $\mathbf{B}_{\mu}$ of radius $p_{F}=\sqrt{\mu^{2}-m^{2}}$ (the Fermi momentum). We thus recover the $B=0$ result of equation~\eqref{eq:Fermi_gas}, up to a $\mu$-independent cosmological constant term coming from $f(B=0,0)$ that we neglect (see eq.~\eqref{eq:weak_mag}). The lowest order correction can be computed explicitly analytically, see appendix~\ref{app:Bweak} for its derivation:
\begin{equation}\label{eq:weak_mu}
\begin{split}
f(B,\mu)\Big\vert_{\rm weak} = -\dfrac{B^2}{2} + f(0,\mu) \,-\dfrac{e^{2}B^{2}}{24 \pi^{2}} \,\arctanh\left(\dfrac{\sqrt{\mu^{2}-m^{2}}}{\mu}\right) + {\rm const} + \mathcal{O}\left((eB)^{5/2}\right).
\end{split}
\end{equation}

\subsection{Magnetic susceptibility and the de Haas--van Alphen effect}

The free energy of the system is not a directly observable quantity. More interesting quantities are the magnetization of the system, or the magnetic susceptibility, describing the response of the system to changes in the background magnetic field. The latter is defined as
\begin{equation}
\chi_B(B,\mu) \equiv - \dfrac{{\rm d}^2  f(B,\mu)}{{\rm d}B^2} -1,
\end{equation}
and describes the \emph{induced} magnetization response.\footnote{Sometimes an alternative definition is used in the literature, in which the classical background contribution ``$1$'' is not subtracted.}
Using eq.~\eqref{eq:fB} we can obtain the magnetic susceptibility of the QED vacuum in the presence of a magnetic background, and an analytic derivation of the de Haas--van Alphen effect.
Since $\chi_B$ is a dimensionless quantity, the result takes a particularly nice form once written in terms of dimensionless variables. The natural variables are $\beta$, measuring the magnetic field strength, and the chemical potential relative to the fermion mass $\tilde{\mu}= \mu/m$. Using
\begin{equation}
\tilde{\mu}\equiv \dfrac{\mu}{m}, \qquad \beta \equiv \dfrac{2 e B}{m^2} , \qquad -\dfrac{\partial^2}{\partial B^2} = - \dfrac{4 e^2}{m^4} \dfrac{\partial^2}{\partial \beta^2},
\end{equation}
the magnetic susceptibility at finite $\mu$ can be expressed as 
\begin{equation}
\label{eq:chiB_full}
\chi_B(B,\mu) = \dfrac{\alpha}{\pi} \dfrac{\partial^2 }{\partial \beta^2} \,g(\beta, \tilde\mu),
\end{equation}
where $\alpha $ is the fine structure constant $\alpha = e^2 / 4\pi$. The function $g(\beta, \tilde\mu)$ is a rescaled version of the quantum contribution to the free energy density~\eqref{eq:full_bmu}, given by\footnote{For simplicity of notation we assume here that $\tilde{\mu}\geq 1$. Notice that $G_0\to 0$ and for $\tilde{\mu}\to 1^+$, so that the finite density contribution vanishes in this limit.}
\begin{equation}
g(\beta,\tilde{\mu}) = \,g(\beta,0) \,+ \dfrac{1}{2} G_0
+ \sum_{\ell =1}^{\ell_{\rm max}} G_\ell,
\end{equation}
where
\begin{equation}\label{eq:Gell}
\begin{split}
&g(\beta,0) = \beta^{2} 
\Bigg(2 \, \zeta' \left(-1,\dfrac{1}{\beta}\right)
+2 \Big(1-\log\beta\Big) \zeta\left(-1,\dfrac{1}{\beta}\right)- \dfrac{1}{\beta}\Bigg), \\
&G_\ell(\beta,\tilde{\mu}) = -2\beta \left((1+\beta \ell)\,\arctanh\left(\dfrac{\sqrt{\tilde{\mu}^2-\beta \ell-1}}{\tilde{\mu}}\right)- \tilde{\mu} \sqrt{\tilde{\mu}^{2}-\beta\ell -1}\right),\\
&\ell_{\rm max} =  \Bigg\lfloor \dfrac{\tilde{\mu}^{2}-1}{\beta}\Bigg\rfloor .
\end{split}
\end{equation}
The last term can be rewritten using the Heaviside theta function as
\begin{equation}
\sum_{\ell =1}^{\infty} G_\ell \;\Theta\left(\ell_{\rm max}-\ell \right),
\end{equation}
and one might worry of picking up $\delta\left(\ell_{\rm max}-\ell \right)$ (and $\delta'$) contributions from the derivatives in eq.~\eqref{eq:chiB_full}. These contributions, however, vanish identically up to second derivatives in~$\beta$, since $G_{\ell}$ and $\partial_\beta G_{\ell}$ vanish identically when $\ell_{\rm max}$ crosses an integer value and $\ell=\ell_{\rm max}$. Noticing also that $G_0$ depends linearly on $\beta$, we arrive at
\begin{equation}\label{eq:chiB_mu}
\chi_B(B,\mu) = \dfrac{\alpha}{\pi} \left( \partial_\beta^2 \,g(\beta, 0) + \sum_{\ell =1}^{\ell_{\rm max}} \partial_\beta^2 G_\ell(\beta,\tilde{\mu}) \right) .
\end{equation}

\begin{figure}[t]
\begin{center}
\includegraphics[width=.7\textwidth]{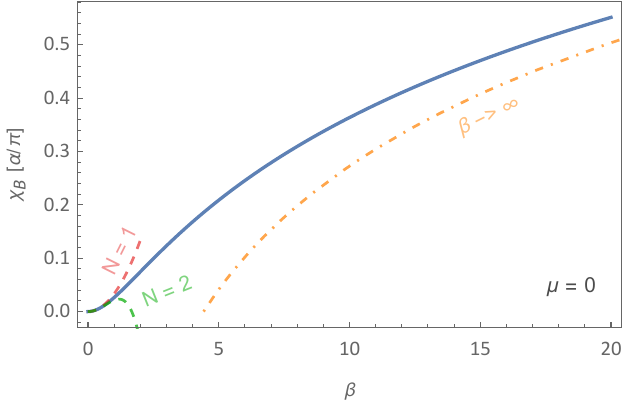}
\vspace{-15pt}
\end{center}
\caption{\label{fig:chiB_vacuum} \it \small Magnetic susceptibility of the zero density QED vacuum in an external magnetic field, in units of $\alpha/\pi$ and as a function of $\beta\equiv 2eB/m^2$. The dashed red ($N=1$) and green ($N=2$) lines provide truncations of the small $\beta$ asymptotic expansion with $N$ terms. Increasing $N$ improves the accuracy for small $\beta$, but reduces the region in which the approximation is valid. The strong field approximation (shown as an orange dot-dashed line) provides an accurate result only for $\beta \gtrsim 100$. The intermediate region is described by the full non-perturbative result.}
\end{figure}

The first term describes the magnetic response of the QED vacuum at finite density, for arbitrary values of the external magnetic field. We display the zero density contribution in figure~\ref{fig:chiB_vacuum}, in units of $\alpha/\pi$. The weak field regime is governed by the asymptotic expansion of the magnetic Euler--Heisenberg Lagrangian~\eqref{eq:weak_mag}. This provides a good approximation only for $\beta <1$. Notice that, being an asymptotic expansion, the region of $\beta$ for which the approximation works well shrinks to zero as we increase the number of terms. In particular, the expansion can never work well beyond $\beta \sim 1$. In the opposite limit of strong field, from eq.~\eqref{eq:strong_mag} we obtain instead 
\begin{equation}
\chi_B\Big\vert_{\rm strong} =\dfrac{\alpha}{\pi}\,  \partial_\beta^2 \,g(\beta, 0)\Bigg\vert_{\beta\to \infty}\simeq \dfrac{\alpha}{\pi}\left(\dfrac{1}{6} + 4\, \zeta'(-1) + \dfrac{1}{3} \log\beta\right).
\end{equation}
The strong field limit provides an estimate accurate at the $10\%$ level for $\beta \gtrsim 20$, and at the $1\%$ level for $\beta \gtrsim 100$. We see, therefore, that in order to describe the magnetic response of zero density QED in the range $1\lesssim \beta \lesssim 20$ the full non-perturbative result is needed.

\begin{figure}
\begin{center}
\includegraphics[width=.7\textwidth]{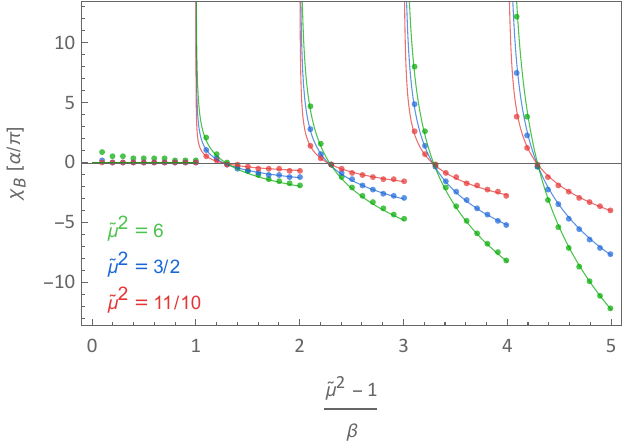}
\vspace{-15pt}
\end{center}
\caption{\label{fig:chiB_mu1} \it \small Magnetic susceptibility of the finite density QED vacuum in a strong external magnetic field, for different values of $\tilde{\mu}= \mu/m$. We use variables that make manifest the filling of the first five Landau levels. The magnetic susceptibility has a periodic spike feature as a function of $1/\beta$, occurring for integer filling, known as the de Haas--van Alphen effect. Continuous lines: analytic form of the finite $\mu$ contribution (second term in eq.~\eqref{eq:chiB_mu}). Dots: full result, evaluated at equidistant points.}
\vspace{20pt}
\end{figure}

Consider now the second term in eq.~\eqref{eq:chiB_mu}, describing the finite density contribution to $\chi_B$. In the limit of very strong magnetic field limit one has $\ell_{\rm max}=0$, so that the magnetic response at finite density is (exactly) the same as that at zero density. For a strong field, but weak enough that at least one Landau level is filled, the finite density contribution dominates. From the expression in eq.~\eqref{eq:Gell} we have
\begin{equation}
\partial_\beta^2 G_\ell(\beta,\tilde{\mu}) =  \frac{\beta  \ell^2 \tilde{\mu}}{(\beta  \ell+1) \sqrt{-\beta  \ell+\tilde{\mu}^2-1}}-4\ell \tanh ^{-1}\left(\frac{\sqrt{-\beta  \ell+\tilde{\mu}^2-1}}{\tilde{\mu}}\right),
\end{equation}
so that the susceptibility diverges whenever $\tilde{\mu}^2-1$ is an integer multiple of $\beta$, and $\ell_{\rm max}$ jumps by one unit. This oscillatory feature, displayed in figure~\ref{fig:chiB_mu1}, occurs every time a Landau level achieves integer filling. This periodic property of the magnetic susceptibility~$\chi_B$ as a function of $\beta$ is known as the de Haas--van Alphen effect~\cite{Landau:1980mil,kittel2005introduction}. The period is given~by
\begin{equation}
\Delta\left( \dfrac{1}{\beta}\right) =  \dfrac{1}{\tilde{\mu}^2-1} \quad \implies \quad \Delta\left( \dfrac{1}{B}\right) =  \dfrac{2e}{\mu^2-m^2} = \dfrac{8\pi e}{A_{F}}, 
\end{equation}
where $A_F= 4\pi p_{F}^2= 4\pi (\mu^2 - m^2)$ is the area of the Fermi surface.

\begin{figure}
\begin{center}
\includegraphics[width=.7\textwidth]{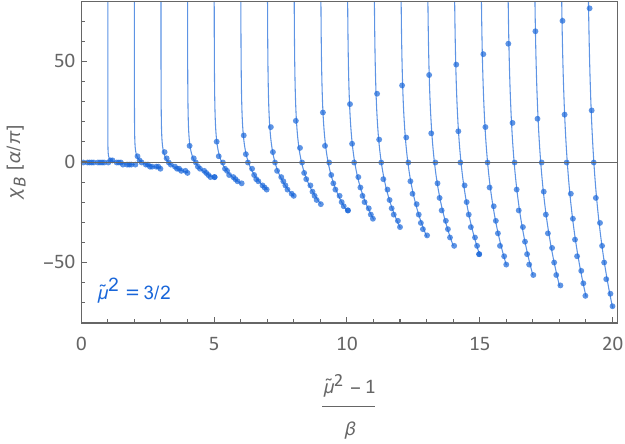}
\vspace{-15pt}
\end{center}
\caption{\label{fig:chiB_mu2} \it \small Same as figure~\ref{fig:chiB_mu1} but for a larger range of $1/\beta$ and for $\tilde{\mu}^2=\mu^2/m^2 = 3/2$. In the weak field limit (small $\beta$), the magnetic susceptibility $\chi_B$ oscillates more and more wildly. The smooth result~\eqref{eq:chiB_weak} is obtained only as an average.}
\end{figure}

In the weak field limit, on the other hand, using eq.~\eqref{eq:weak_mu} we obtain the continuous approximation for the susceptibility
\begin{equation}\label{eq:chiB_weak}
\chi_B\Big\vert_{\rm weak} = \dfrac{\alpha}{3\pi} \,\arctanh \left( \dfrac{\sqrt{\tilde{\mu}^2-1}}{\tilde{\mu}}\right) + \mathcal{O}\left(\sqrt{B} \right) .
\end{equation}
As we show in figure~\ref{fig:chiB_mu2}, the weak field approximation is obtained only as an average.


\section{The four dimensional Gross--Neveu model at finite density}
\label{sec:interacting}

Having understood the case of free fermions in the path integral approach, we would like to include interaction terms among the fermions and see if our conclusions are modified by the presence of interactions.
To this end, we shall consider a generalization of the Gross--Neveu model~\cite{Gross:1974jv} to $d=3+1$ dimensions, and work in the $1/N$ expansion.\footnote{The dynamics of the Gross--Neveu model in $2+1$ dimensions has received renewed attention in connection to lattice investigations. Its behavior at finite density and in the presence of a magnetic background has been studied in~\cite{Lenz:2023gsq}.}

In our analysis we make the assumption that spatial translational invariance is unbroken by the ground state (i.e., that the system is in a homogeneous phase). Spatially inhomogeneous phases are known to occur in the Gross--Neveu model in $1+1$ dimensions~\cite{Thies:2006ti} and in its chiral version~\cite{Ciccone:2022zkg}, however there is some evidence~\cite{Pannullo:2023one} that homogeneous phases are stable in $2+1$ dimensions. We shall assume this to be the case also in $3+1$ dimensions. 

In $3+1$ dimensions, the Gross--Neveu model can be defined for a system of $N$ Dirac fermions transforming in the fundamental representation of a (vectorial) $\U(N)$ global symmetry group. We consider the massless case for simplicity, but our discussion can be extended to include a mass term. The interaction term is a four-fermion interaction:
\begin{equation}
\mathcal{L}= \bar{\Psi}_a \( i \slashed{\partial} \) \Psi_a - \dfrac{\lambda}{2 N} \left(\bar{\Psi}_a \Psi_a \right)^2,
\end{equation}
where the sum over the flavor index $a$ is implicit. This action has a $\U(N)$ symmetry of which $\U(1)_V$ is a subgroup. The axial symmetry $\U(1)_A$ is explicitly broken by the interaction term, however a discrete axial $\mathbb{Z}_2^A$ survives, acting on the fermions as:
\begin{equation}
\Psi_a \longrightarrow \gamma_5\Psi_a.
\end{equation}
This symmetry forbids the mass term, and prevents its appearance under renormalization group flow.

\subsection{Zero density}

We can rewrite the Lagrangian as a quadratic fermionic Lagrangian by introducing a scalar auxiliary field~\cite{Gross:1974jv}. To do so we simply add a Gaussian term to the Lagrangian:
\begin{equation}
\delta \mathcal{L} = \dfrac{N}{2\lambda}\left(\sigma -  \dfrac{\lambda}{N} \bar{\Psi}_a \Psi_a \right)^2,
\end{equation}
which has the only effect of changing the normalization of the path integral by a constant factor.
The four-fermion interaction terms are canceled by the auxiliary term, and we are left with 
\begin{equation}
\mathcal{L} = \bar{\Psi}_a \( i \slashed{\partial} - \sigma \) \Psi_a + \dfrac{N}{2\lambda} \sigma^2 .
\end{equation}
The discrete axial symmetry $\mathbb{Z}_2^A$ acts on $\sigma$ as 
\begin{equation}
\sigma  \longrightarrow - \sigma.
\end{equation}
In the large $N$ limit, $N\rightarrow \infty$ with $\lambda$ fixed, the auxiliary field $\sigma$ becomes infinitely heavy, and the $\sigma$ propagator is suppressed by a factor $1/N$. At leading order in the $1/N$ expansion we can therefore treat the scalar field $\sigma$ at tree level, and consider fermionic loops only. 

The Gross--Neveu model is renormalizable in two dimensions, and no additional counterterms are needed beyond the four-fermion interaction, or equivalently the quadratic term for the auxiliary field $\sigma$. In four dimensions, on the other hand, the theory is non-renormalizable, and an infinite number of counterterms would be needed to renormalize the theory to all orders. The situation simplifies however at leading order in large $N$: a single counterterm $g \sigma^4\,$ is needed to renormalize the theory, to all orders in the coupling~$\lambda$.

To see this, consider now $Z[\sigma;0]$, which can be computed by standard methods, analytically continuing the momenta to Euclidean space. Up to an additive constant, we have: 
\begin{equation} 
V_{\rm eff}(\sigma)= i \,\log Z[\sigma;0] = - \dfrac{4N}{2} \int \dfrac{{\rm d}^{d}p}{(2\pi)^{d}} \log \left(p^2+\sigma^2\right) -  \dfrac{N}{2\lambda} \sigma^2 ,
\label{VeffGN}
\end{equation}
where $p$ is in Euclidean space, i.e.~we replaced $p^0\mapsto i p^0$.
We regulate the theory adopting dimensional regularization. The divergent contribution for $d\rightarrow 4$ is
\begin{equation}
V_{\rm eff}(\sigma) \Big\vert_{\rm div} =-\dfrac{N}{8\pi^2} \left(\dfrac{1}{d-4}\right)\,\sigma^4.
\end{equation}
We can renormalize  the theory by including of a local $g_4 \sigma^4\,$ counterterm in the action. In the $\overline{\rm MS}$ scheme, neglecting the cosmological constant term, we obtain
\begin{equation}
V_{\rm eff}(\sigma)= \dfrac{N}{32\pi^2} \left(3-2\log \left(\dfrac{\sigma^2}{{\bar{\mu}}^2}\right) \right) \sigma^4 -  \dfrac{N}{2\lambda} \sigma^2 +  \bar{g}_4 \,\sigma^4,
\label{VeffGN2}
\end{equation}
where $\bar{\mu}$ denotes the renormalization scale and $\bar{g}_4$ is the $\overline{\rm MS}$ running coupling. The effective potential, as it stands, has an unstable direction for large values of $\sigma $, due to the negative logarithmic contribution enhanced by the large $N$ factor. We are therefore forced to take into account higher dimensional operators, that stabilize the theory. For this purpose we include higher order terms in the auxiliary scalar field formulation, taking the Lagrangian
\begin{equation}
\mathcal{L} = \bar{\Psi}_a \( i \slashed{\partial} - \sigma \) \Psi_a + \dfrac{N}{2\lambda} \sigma^2 + g_4 \,\sigma^4 + g_6 \,\sigma^6 + \dots ,
\label{LGNmu0}
\end{equation}
as the \emph{definition} of the fermionic model. Only even powers of $\sigma$ appear, to preserve the discrete axial symmetry $\mathbb{Z}_2^A$.
For the purpose of illustration, we consider the case $g_6>0$ and $g_{2n}=0$ for $n>3$, but the results we discuss can be generalized straightforwardly. In the purely fermionic formulation this corresponds to adding an infinite series of local interactions of the form $(\bar{\Psi}_a \Psi_a)^n$, with coefficients determined by $\lambda,g_4,g_6$ and $N$.

At leading order in $1/N$, the coupling $\lambda$ does not run and the large $N$ theory can be defined by specifying the (arbitrary) value of $\lambda$. 
We can then express the running quartic coupling $\bar{g}_4(\bar{\mu})$ in terms of a dimensionless physical coupling $\kappa_4$, evaluated at the scale 
\begin{equation}
M \equiv \dfrac{1}{\sqrt{\vert\lambda \vert}}.
\end{equation}
Computing the beta function of $\bar{g}_4(\bar{\mu})$, and solving its RG evolution, we find
\begin{equation}
{\bar{\mu}}\dfrac{\d}{\d {\bar{\mu}}}\bar{g}_4 = - \dfrac{N}{8\pi^2}\quad \implies\qquad \bar{g}_4 = \kappa_4 - \dfrac{N}{16\pi^2} \log \left(\dfrac{{\bar\mu}^2}{M^2}\right).
\end{equation}
Similarly, the coupling $g_6$ can be expressed in terms of a dimensionless coupling $\kappa_6$ by defining
\begin{equation}
g_6 = \dfrac{\kappa_6}{M^2}.
\end{equation}
We shall assume that $M$ is the only dimensionful scale in the problem, so that $\kappa_4, \kappa_6$ are of order one. It is convenient to adopt units $M=1$ (factors of $M$ can always be reintroduced by dimensional analysis), and set $\kappa_2 \equiv {\rm sign}(\lambda)$.

Expressing the effective potential in terms of physical couplings we arrive at
\begin{equation}\label{eq:GNeffpotential}
V_{\rm eff}(\sigma)= \dfrac{N}{32\pi^2} \left(3-2\log \sigma^2 \right) \sigma^4 - \dfrac{N}{2} \kappa_2 \, \sigma^2 +  \kappa_4 \, \sigma^4 + \kappa_6 \, \sigma^6.
\end{equation}
In taking the large $N$ limit we keep the couplings $\kappa_4, \kappa_6$ fixed. Loops involving $\kappa_4$ and $\kappa_6$ vertices, and $\sigma$ as an internal line, are therefore subleading in the $1/N$ expansion.
We notice that the $\bar{\mu}$ dependence of the effective potential drops out once we express the result in terms of physical couplings. This property is consistent with the fact that the minimum of the effective potential is related to observable quantities.

The effective potential~\eqref{eq:GNeffpotential} at large $N$ has a global minimum $\sigma_0$ at  
\begin{equation}
\sigma^2_0 = \dfrac{N}{24\pi^2 \kappa_6}\left(\log\left(\dfrac{N}{24\pi^2 \kappa_6}\right) + \mathcal{O}\left(\log \log N \right) \right). 
\end{equation}
A more accurate estimate can be obtained in terms of the negative branch of the Lambert $W$ function (neglecting $\kappa_2,\kappa_4$, which are suppressed at large values of $\sigma_0$)
\begin{equation}
\sigma^2_0 = \dfrac{W_{-1}(x)}{x} \e, \qquad x = \dfrac{24\pi^2 \e \kappa_6}{N}.
\end{equation}
This minimum always exists for $\lambda>0$ ($\kappa_2=+1$) and $\kappa_{4},\kappa_{6}\approx 1$ , whereas it requires large enough $N$ for $\lambda<0$ ($\kappa_2=-1$). The large $N$ scaling of $\sigma_0$ would be modified by the presence of higher order interactions, e.g.~$\kappa_8\approx 1$, however the qualitative features of the model would be unchanged.

The expectation value of $\sigma$ breaks spontaneously the discrete $\mathbb{Z}_2^A$ symmetry, generating an effective mass term $m_{\rm eff}=\sigma_0$ for the fermions, similar to the mass generation mechanism for quarks and leptons by the Higgs field in the Standard Model. The vectorial symmetry $\U(N)$, and in particular $\U(1)_V$, is instead unbroken.

\subsection{Finite density}
We can introduce a chemical potential for the $\U(1)_V \subset \U(N)$ symmetry, in such a way that the $\SU(N)$ subgroup is preserved:
\begin{equation}
\mathcal{L}_{\mu} = \bar{\Psi}_a \( i \slashed{\partial} + \gamma^{0} \mu - \sigma \) \Psi_a + \dfrac{N}{2\lambda} \sigma^2 + g_4 \,\sigma^4 + g_6 \,\sigma^6 .
\end{equation}
We would like to show that this theory can support a finite density phase with unbroken $\U(1)_V$ symmetry. A necessary condition for this is that the ground state energy density of the system has a non-trivial $\mu$ dependence, for $\mu$ larger than some critical value $\mu_{\rm crit}$. We will show that in the large $N$ fermionic model under consideration this expectation can be met without the need for $\U(1)_V$ breaking order parameters. This behavior should be contrasted with that observed in the case of interacting scalar fields, and in particular in the ${\rm O}(N)$ model with quartic interactions, analyzed in~\cite{Nicolis:2023pye}, where it is shown that the system can support a finite density phase only in the presence of a non-zero expectation value for an $\U(1)_V$ breaking order parameter.
We shall find that the critical value $\mu_{\rm crit}$ to support finite density coincides with $\sigma_0$, the effective (pole) mass of the lowest lying charged states, as expected on physical grounds and argued also in the scalar case~\cite{Nicolis:2023pye}.

In the background of $\sigma$, the fermionic part of the action at finite $\mu$ is simply given by
\begin{equation}\label{eq:GNfermionmu}
\mathcal{L}_{\mu}\Big\vert_{\Psi} = \bar{\Psi}_a \( i \slashed{\partial} + \gamma^{0} \mu - \sigma \) \Psi_a.
\end{equation}
This amounts to $N$ copies of the free fermion~\eqref{eq:free}, with $m\rightarrow \sigma$. The finite $\mu$ fermionic path integral can therefore be computed exactly using the contour integration method previously described.
We are interested in computing
\begin{equation}\label{eq:Veffsm}
V_{\rm eff}(\sigma;\mu) \equiv i \left(\log \dfrac{Z[\sigma;\mu]}{Z[0,0]}\right).
\end{equation}
From this we will compute the free energy density of the system by setting the auxiliary field at its background value:
\begin{equation}
f(\mu)= V_{\rm eff}(\sigma_{\rm min}(\mu);\mu), \qquad \qquad \dfrac{{\rm d }V_{\rm eff}}{{\rm d}\sigma} (\sigma_{\rm min}(\mu);\mu)=0.
\end{equation}
We already computed  $\mu$ effective potential $V_{\rm eff}(\sigma;0)$ in eq.~\eqref{eq:GNeffpotential}. All we need to compute is, therefore, the difference $V_{\rm eff}(\sigma;\mu) - V_{\rm eff}(\sigma;0) $. 
Given the fermionic action~\eqref{eq:GNfermionmu}, the contribution $V_{\rm eff}(\sigma;\mu) - V_{\rm eff}(\sigma;0)$ can be computed following the same complex contour integration technique we used for free fermions in the path integral formalism, as in sec.~\ref{subsec:free}. 
The result is simply given by $N$ times~\eqref{eq:ffree}, after the substitution $m\rightarrow \sigma$:
\begin{equation}\label{eq:Vmu}
\begin{split}
V_\mu(\sigma) &\equiv V_{\rm eff}(\sigma;\mu) - V_{\rm eff}(\sigma;0) \\
&= \Theta\Big(\mu^2-\sigma^2\Big) \dfrac{N}{24\pi^{2}}\left(  \mu (5\sigma^{2}-2\mu^{2})\sqrt{\mu^{2}-\sigma^{2}} - 3 \sigma^{4} \arctanh\left(\dfrac{\sqrt{\mu^{2}-\sigma^{2}}}{\mu}\right) \right).
 \end{split}
\end{equation}

It follows immediately that as long as $\mu< \sigma_0$ the effective potential is not modified in a neighborhood of $\sigma_0$, the expectation value of $\sigma$ at $\mu=0$. One can easily check that for $\mu< \sigma_0$ the global minimum of the potential $V_{\rm eff}(\sigma;\mu)$ is still given by $\sigma_0$. The fermions have zero charge density, since
\begin{equation}\label{eq:chargevev}
Q(\mu) \equiv \langle   \hat{J}^{0}   \rangle_{\mu} =-i \dfrac{{\rm d} }{{\rm d} \mu}\log Z(\mu)= -\dfrac{{\rm d} V_{\rm eff}(\sigma_{\rm min}(\mu);\mu)}{{\rm d}\mu},
\end{equation}
and
\begin{equation}
Q(\mu)=  -\dfrac{{\rm d} V_{\rm eff}(\sigma_0;\mu)}{{\rm d}\mu} =0 , \qquad\qquad {\rm for} \qquad \mu< \sigma_0.
\end{equation}

For $\mu \geq  \sigma_0$ the situation changes, and the value $\sigma_0(\mu)$ at which the potential $V_{\rm eff}(\sigma;\mu)$ is minimized acquires a non-trivial $\mu$ dependence. Therefore $\mu_{\rm crit}=\sigma_0$, the pole mass of the fermionic excitations in the zero-density theory. The value of the effective potential at the minimum acquires itself a $\mu$ dependence, so that the system supports a finite charge density. Let us start analyzing the small density limit. For $\mu>\sigma_0$ and $\delta\mu\equiv\mu - \sigma_0 \ll \sigma_0$ we find that the effective potential in a neighborhood of $\sigma_0$ is given by 
\begin{equation}
\begin{split}
V_{\rm eff}(\sigma;\mu) = &- \dfrac{N}{96\pi^2}\left(-5+4 \log\sigma_0 \right)\sigma_0^4 +
 \dfrac{N\sqrt{8}}{3\pi^2} (\delta\mu\cdot \sigma_0)^{3/2} \,(\sigma-\sigma_0)\\
&+\left( 6 \kappa_6\, \sigma_0^2 + \dfrac{3N}{4\sqrt{2}\pi^2} (\delta\mu^3 \cdot \sigma_0)^{1/2} -  \dfrac{N}{\sqrt{2}\,\pi^2} (\delta\mu \cdot \sigma_0^3)^{1/2} \right)\,(\sigma-\sigma_0)^2 + \dots
\end{split}
\end{equation}
The potential is now minimized at 
\begin{equation}
\sigma_{\rm min}(\mu)= \sigma_0  - \dfrac{N}{9\sqrt{2} \, \pi^2\kappa_6} \left(\dfrac{\delta\mu}{\sigma_0}\right)^{3/2}\sigma_0  - \dfrac{N^2}{108 \, \pi^4 \kappa_6} \left(\dfrac{\delta\mu}{\sigma_0}\right)^{2}\sigma_0+\dots,
\end{equation}
so that the free energy density is given by 
\begin{equation}
f(\mu)=  - \dfrac{N}{96\pi^2}\left(-5+4 \log\sigma_0 \right)\sigma_0^4 -  \dfrac{N^2}{27 \,\pi^4\, \kappa_6} \delta\mu^3 \, \sigma_0 - \dfrac{N^3}{162\sqrt{2} \,\pi^6\, \kappa_6^2}  \delta\mu^{7/2} \, \sqrt{\sigma_0}\dots 
\end{equation}
Correspondingly, the charge density of fermions is given by
\begin{equation}
Q(\mu)=  -\dfrac{{\rm d} V_{\rm eff}(\sigma_{\rm min}(\mu);\mu)}{{\rm d}\mu}= \dfrac{N^2}{9 \,\pi^4\, \kappa_6} \delta\mu^2 \, \sigma_0 + \dfrac{7 N^3}{324\sqrt{2} \,\pi^6\, \kappa_6^2}  \delta\mu^{5/2} \, \sqrt{\sigma_0}+\dots 
\end{equation}

\begin{figure}
\begin{center}
\includegraphics[width=4in]{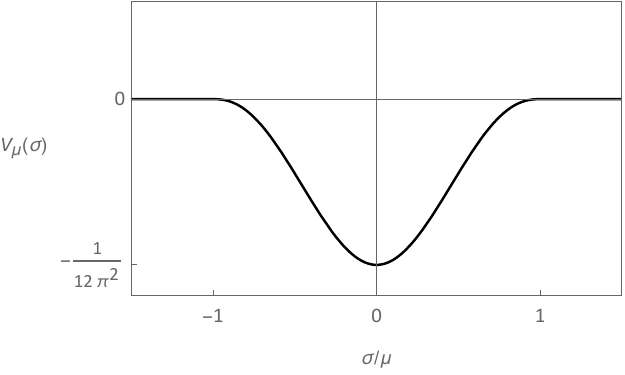}
\vspace{-15pt}
\end{center}
\caption{\label{fig:Vmu} \it \small Finite $\mu$ contribution to the effective potential of $\sigma$ in the (generalized) Gross--Neveu model in $3+1$ dimensions, eq.~\eqref{eq:Vmu}. The result is normalized in units of $N \mu^4$ and expressed as a function of~$\sigma/\mu$.}
\end{figure}

\begin{figure}
\begin{center}
\includegraphics[width=4in]{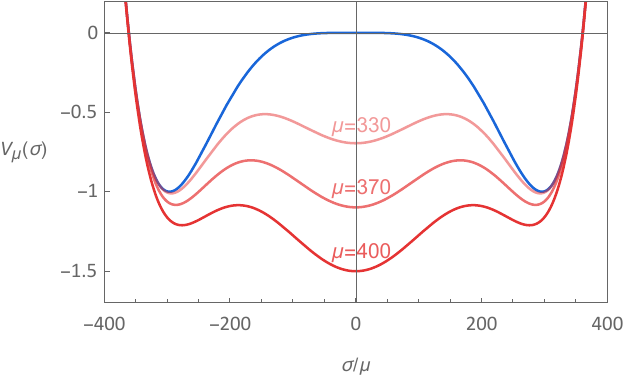}
\vspace{-15pt}
\end{center}
\caption{\label{fig:Vtot} \it \small In blue: effective potential for a (generalized) Gross--Neveu model, in units $M=1$, for $\kappa_2=1,\kappa_4=2,\kappa_6=0.5$ and $N=10^6$, see eq.~\eqref{eq:GNeffpotential}. For our choice of parameters, the $\mu=0$ minimum is located at $\sigma_0 \approx 296$, and we normalized the potential in such a way that its value at $\sigma_0$ is~$-1$. In red: finite $\mu$ effective potential $V_{\rm eff}(\sigma;\mu)$, for increasing values of $\mu= 330$, $\mu = 370$ and $\mu=400$ (shown with growing intensity). There is a $\mathbb{Z}_2^A$ symmetry restoring first order phase transition at a critical value of $\mu$, approximately given by $\mu_A\approx 370$. For our choice of parameters, at the phase transition the charge density jumps from $Q_{-}(\mu_A)\approx 0.25$ to $Q_{+}(\mu_A)= 1$ in units of $\frac{N}{3\pi^2} \mu_A^3$, see~eq.~\eqref{eq:GNrestoredQ}.}
\end{figure}

For small charge densities, we see that the discrete axial symmetry $\mathbb{Z}_2^A$ is in the broken phase. The symmetry breaking scale $\sigma_{\rm min}(\mu)$ decreases with increasing chemical potential and, correspondingly, charge density.
This behavior can be understood by noticing that the finite $\mu$ contribution to $V_{\rm eff}$ is a monotonic increasing function of $\sigma$, for positive $\sigma$, starting from a value of $-\frac{N}{12\pi^2}\mu^4$ for $\sigma=0$ and increasing up to zero for $\sigma\geq \mu$ (see figure~\ref{fig:Vmu}). 
As a consequence, in the limit of large $\mu$ the $\mathbb{Z}_2^A$ symmetry will be restored, since the minimum of the effective potential $V_{\rm eff}(\sigma;\mu)$ will be in zero. This phase transition is in fact of \emph{first order} and has an associated \emph{latent charge density}, as can be readily verified by studying the full closed form effective potential we derived. A system with a charge density falling in between these two values is in a mixed phase and has chemical potential $\mu_A$. We provide a representative plot of the symmetry restoration phase transition in figure~\ref{fig:Vtot}.

An analytic estimate for the critical value of chemical potential $\mu_{A}$ corresponding to the $\mathbb{Z}_2^A$ symmetry restoration phase transition can be derived by comparing the value of the effective potential at the $\sigma=0$ minimum, with the value of the $\mu=0$ effective potential at the minimum $\sigma =\sigma_0$. Since the finite $\mu $ contribution is always non-positive, this estimate provides a lower bound on $\mu_{\rm A}$. We find
\begin{equation}
V_{\rm eff}(0;\mu) = -\dfrac{N}{12\pi^2} \mu^4, \quad V_{\rm eff}(\sigma_0;0) \simeq  - \dfrac{N}{96\pi^2}\left(-5+4 \log\sigma_0 \right)\sigma_0^4,
\end{equation}
so that
\begin{equation}
\mu_A  \gtrsim  \left(-\dfrac{5}{8}+\dfrac{1}{2} \log\sigma_0 \right)^{1/4} \sigma_0,
\end{equation}
where in our estimate we have assumed $\sigma_0$, and therefore $N$, to be large enough. The symmetry restoration chemical potential $\mu_A$ is parametrically close to the critical chemical potential necessary to have a non-zero charge density, $\mu_{\rm crit}=\sigma_0$, with their ratio growing with $\sigma_0$ in a very mild way.

For $\mu>\mu_A$, the symmetry $\mathbb{Z}_2^A$ is restored, so that the fermionic excitations become effectively massless, and the properties of the system become equivalent to those of a system of $N$ free massless Dirac fermions. Perhaps counterintuitively, the effect of the interactions induced by $\lambda, g_4, g_6$ is overcome by the finite density contribution, driving the system to an effectively \emph{free} finite density phase:
\begin{equation}
f(\mu)= V_{\rm eff}(\sigma_{\rm min}(\mu);\mu) =- \dfrac{N}{12\pi^2} \mu^4 , \qquad\qquad {\rm for} \qquad \mu > \mu_A.
\end{equation}
The charge density, in particular, is simply given by 
\begin{equation}\label{eq:GNrestoredQ}
Q(\mu)=  -\dfrac{{\rm d} V_{\rm eff}(\sigma_{\rm min}(\mu);\mu)}{{\rm d}\mu}=  \dfrac{N}{3\pi^2} \mu^3 , \qquad\qquad {\rm for} \qquad \mu > \mu_A.
\end{equation}
We stress that this conclusion has general validity in this class of large $N$ models, and holds even with the inclusion of higher order polynomial interactions for $\sigma$.

Before concluding this section, let us comment on the question we started from: the possibility of $\U(1)_V$ symmetry breaking at finite density. Our analysis of the Gross--Neveu model does not formally exclude this possibility, however it provides partial evidence for the hypothesis that the $\U(1)_V$ symmetry remains unbroken. Indeed, the finite $\mu$ free energy has a non-trivial $\mu$ dependence and the system can support a finite charge density with unbroken $\U(1)_V$, differently from the scalar case analyzed in~\cite{Nicolis:2023pye}, where  the breaking of $\U(1)_V$ was shown to be necessary to support finite density in an ${\rm O}(N)$ model. Moreover, the value of $\mu_{\rm crit}$ that we found coincides with the fermion pole mass, as expected on physical grounds, further hinting at the consistency of the scenario. 
Note that the argument discussed in sec.~\ref{sec:jj_term} for Majorana order parameter can not be directly extended to the large $N$ model, due to the $\U(N)$ vectorial symmetry. Indeed, every Majorana-like fermion bilinear would transform in a non-trivial representation of the $\SU(N)$ subgroup of $\U(N)$, and cannot be regarded as an order parameter for $\U(1)_V$ only. The simplest $\U(1)_V$ scalar order parameter can be constructed, for $N$ even, contracting the indices of $N$ fermions with a completely antisymmetric tensor, and is never a fermion bilinear for $N>2$. 

\section{Discussion and outlook}

The dynamics of interacting systems of fermions at finite density and low temperature is far from being fully understood. A common belief is that, barring the case of free fermions (which form a Fermi gas), interacting fermions at low temperatures flow to a superfluid phase, or possibly to a non-Fermi liquid phase. The argument for this expected behavior relies on the existence of relevant deformations in the EFT of Fermi liquids~\cite{Benfatto:1990zz,Polchinski:1992ed,Shankar:1993pf},\footnote{See also the interesting recent work~\cite{Delacretaz:2022ocm} for an alternative approach.} that can generically lead to the formation of Cooper pairs, or more exotic strong coupling behavior. To the best of our knowledge, however, a complete understanding of the IR phases of interacting fermions is still missing, especially in the relativistic case, and the question of whether an interacting Fermi \emph{liquid} can exist as a stable zero-temperature phase appears to be an open question.

In this work we described an approach to treat fermionic QFTs at finite chemical potential and zero temperature using path integral techniques. The method relies on an accurate treatment of the $i\varepsilon$ term needed to project on the ground state, and allows to compute finite $\mu$ quantities such as the free energy of a Fermi gas. We leveraged this technical tool to compute analytically the free energy of finite density QED in a homogeneous magnetic background, generalizing the Euler--Heisenberg effective action to finite density and reproducing the de Haas--van Alphen effect. These findings can be of interest in astrophysical contexts, such as strongly magnetized pulsars and relativistic plasmas.

As an application, we studied a generalization of the Gross--Neveu model to $3+1$ spacetime dimensions, and showed that in the large $N$ limit it can support a finite density phase with unbroken $\U(1)_V$ internal symmetry. This is to be contrasted with the case of the scalar ${\rm O}(N)$ vector model, where a finite  density phase can only exist along with the spontaneous breaking of the $\U(1)$ charge associated to $\mu$~\cite{Nicolis:2023pye}, leading to a superfluid behavior. This suggests that there might exist theories of interacting fermions, in physical spacetime dimensions, with a zero-temperature Fermi liquid
phase. A definite answer requires a stability analysis, together with a study of $1/N$ corrections and the dynamics of excitations. We hope to address these questions in the future, and make further progress towards the goal of a general understanding of the infrared phases of relativistic quantum fields at finite density.

\acknowledgments

It is a pleasure to thank Alberto Nicolis for collaboration at the early stages of this work and for useful discussions. We also thank Austin Joyce for discussions and collaboration on related topics. LS is supported by the French Centre National de la Recherche Scientifique (CNRS). AP is supported by the DOE grant DE-SC0011941.

\appendix
\section{Conventions on gamma matrices}
\label{app:conventions}

We adopt the chiral basis for the $\gamma$ matrices
\begin{equation}
\gamma^0 = 
\begin{pmatrix}
0 &  \mathbb{1} \\
\mathbb{1} & 0
\end{pmatrix},
\quad
\gamma^i = 
\begin{pmatrix}
0 &  \sigma^{i} \\
-\sigma^{i} & 0
\end{pmatrix}.
\end{equation}
Weinberg's convention amounts to the replacement
\begin{equation}
\eta_{\mu\nu} \rightarrow -\eta_{\mu\nu}=
\begin{pmatrix}
-1 &  0 & 0 & 0\\
0 &  +1 & 0 & 0\\
0 &  0 & +1 & 0\\
0 &  0 & 0 & +1
\end{pmatrix},\\
\qquad
\gamma^{\mu} \rightarrow i \gamma^{\mu}.
\end{equation}
Charge conjugation is given by:
\begin{equation}
\Psi^{c} = -i \gamma^{2} \Psi^{\star}.
\end{equation}
The charge conjugation matrix is defined by the condition 
\begin{equation}
C \gamma^{\mu} C^{-1}= - (\gamma^{\mu})^{T}, 
\end{equation}
and satisfies the properties
\begin{equation}
C^2 = -1, \qquad C^{-1} = C^{T} = C^{\dagger}= -C, \qquad C= -i \gamma^{0} \gamma^{2}= \begin{pmatrix}
i \sigma^{2} &  0 \\
0 & -i\sigma^{2}
\end{pmatrix}.
\end{equation}

\section{Fermi gas in a magnetic field: weak field expansion}
\label{app:Bweak}
To include $\mathcal{O}(eB)$ corrections in the weak field limit we need an accurate approximation of the sum as an integral. To do this we use the Euler--MacLaurin formula
\begin{equation}
\sum_{\ell=m}^{n} F(\ell) - \int_{m}^{n}F(\ell) {\rm d}\ell = \dfrac{F(m)+F(n)}{2} + \dfrac{F'(n)-F'(m)}{12} + \int_{m}^{n}F'''(\ell) \dfrac{B_{3}(\ell-\lfloor \ell \rfloor)}{3 !} {\rm d}\ell ,
\end{equation}
to approximate equation~\eqref{eq:full_bmu}, considering the sum from $m=0$ to $n=\ell_{\rm max}-1$. We obtain
\begin{align}
\nonumber f(B,\mu)\Big\vert_{\rm weak} = \,&f(B,0) +\dfrac{1}{4\pi^{2}} \int_{0}^{q^{2}_{\rm max}} {\rm d}q^{2} \int_{0}^{\sqrt{\mu^{2}-m^{2}-q^{2}}} {\rm d}p_{z}  \left( 
\sqrt{ (p_{z})^{2}+q^{2}+m^{2}} - \mu \right) \\
&- \dfrac{e^{2}B^{2}}{24 \pi^{2}} \arctanh\left(\dfrac{\sqrt{\mu^{2}-m^{2}}}{\mu}\right) +\mathcal{O}\left((eB)^{5/2}\right) ,
\end{align}
where $q^{2}_{\rm max} =2eB (\ell_{\rm max}-1)$, and we used the fact the $F(\ell_{\rm max})\sim \mathcal{O}\left((eB)^{5/2}\right)$ and $F^{(p)}(\ell_{\rm max}-1) \sim \mathcal{O}\left((eB)^{5/2}\right)$ for $p=0,1,2$.
One can now recognize the integral as an integral in cylindrical coordinates over a domain $\mathbf{\Omega}$ which is the intersection of a ball $\mathbf{B}_{\mu}$ of radius $\sqrt{\mu^{2}-m^{2}}$ and a concentric cylinder $\mathbf{C}_{B}$ of infinite length and radius $q_{\rm max}$. The boundary of $\mathbf{\Omega}$ is the Fermi surface $\partial \mathbf{\Omega}$. We can write the free energy as
\begin{equation}
\begin{split}
f(B,\mu)\Big\vert_{\rm weak} = \,&f(B,0) \,-\dfrac{e^{2}B^{2}}{24 \pi^{2}} \arctanh\left(\dfrac{\sqrt{\mu^{2}-m^{2}}}{\mu}\right) \\
&+2 \int_{\mathbf{\Omega}} \dfrac{{\rm d}^{3}\vec{p}}{(2\pi)^{3}} \left( \sqrt{\vec{p}^{\, 2}+m^{2}} - \mu \right)+\mathcal{O}\left((eB)^{5/2}\right).
\end{split}
\end{equation}
In the limit $2eB \rightarrow 0$ the radius of the cylinder tends to $q_{\rm max}=\sqrt{\mu^{2}-m^{2}}$ and $\mathbf{\Omega}$ tends to the ball $\mathbf{B}_{\mu}$. We thus recover the $B=0$ result of equation~\eqref{eq:Fermi_gas}, up to a $\mu$-independent cosmological constant term coming from $f(B=0,0)$ that we drop. For small non-zero magnetic field we approximate the sum as an integral and write the result as a correction to the $B=0$ case
\begin{equation}\label{eq:weak0}
\begin{split}
f(B,\mu)\Big\vert_{\rm weak} = \,&f(B,0) +f(0,\mu) \,-\dfrac{e^{2}B^{2}}{24 \pi^{2}} \arctanh\left(\dfrac{\sqrt{\mu^{2}-m^{2}}}{\mu}\right) \\
&-2 \int_{\mathbf{B}_{\mu} \setminus \mathbf{\Omega}} \dfrac{{\rm d}^{3}\vec{p}}{(2\pi)^{3}} \left( \sqrt{\vec{p}^{\, 2}+m^{2}} - \mu \right)+\mathcal{O}\left((eB)^{5/2}\right).
\end{split}
\end{equation}
The last integral can be computed in closed form 
\begin{equation}
\int_{\mathbf{B}_{\mu} \setminus \mathbf{\Omega}} \dfrac{{\rm d}^{3}\vec{p}}{(2\pi)^{3}} \left( \sqrt{\vec{p}^{\, 2}+m^{2}} - \mu \right) = \dfrac{1}{96 \pi ^2} \left( 3 \left(\Delta ^2-\mu ^2\right)^2 \arctanh\left(\frac{\Delta }{\mu }\right)+5 \Delta ^3 \mu -3 \Delta  \mu ^3 \right),
\end{equation}
where $\Delta= \sqrt{\mu^{2}-m^{2}-q_{\rm max}^{2}}$ is a small parameter in the limit $2eB \rightarrow 0$, of order \\$\Delta~\sim~\mathcal{O}(\sqrt{eB})$. We can expand this result in powers of $\Delta$ obtaining
\begin{equation}
\dfrac{1}{30 \pi ^2} \left(\frac{\Delta ^5}{\mu }+\frac{\Delta ^7}{7 \mu ^3}+\dots \right).
\end{equation}
Combining this information with eqs.~\eqref{eq:weak0},~\eqref{eq:fB} we obtain the weak field approximation
\begin{equation}
f(B,\mu)\Big\vert_{\rm weak} = -\dfrac{B^2}{2} + f(0,\mu) \,-\dfrac{e^{2}B^{2}}{24 \pi^{2}} \arctanh\left(\dfrac{\sqrt{\mu^{2}-m^{2}}}{\mu}\right) + {\rm const} + \mathcal{O}\left((eB)^{5/2}\right).
\end{equation}

\bibliographystyle{JHEP}  
\bibliography{biblio.bib}

\end{document}